\documentclass[journal]{IEEEtran}
\usepackage[pdftex]{graphicx}
\usepackage{amsmath}
\usepackage{amssymb}
\usepackage{blindtext}
\usepackage{color}
\usepackage{epstopdf}

\ifCLASSINFOpdf
\else
\fi

\begin{document}
\title{Directive Surface Plasmons on Tunable Two-Dimensional Hyperbolic Metasurfaces and Black Phosphorus: Green's Function and Complex Plane Analysis}


\author{\IEEEauthorblockN{S. Ali Hassani Gangaraj~\IEEEmembership{Student member,~IEEE}, Tony Low, Andrei Nemilentsau, George W. Hanson,~\IEEEmembership{Fellow,~IEEE}}

\thanks{Manuscript received xxx; revised xxx. 

Corresponding author: S. Ali Hassani Gangaraj is with Department of Electrical Engineering, University of Wisconsin-Milwaukee, Milwaukee, Wisconsin 53211, USA. (email: ali.gangaraj@gmail.com).

Tony Low is with Department of Electrical and Computer Engineering, University of Minnesota, Minneapolis, Minnesota 55455, USA. (email: tlow@umn.edu).

Andrei Nemilentsau is with Department of Electrical Engineering, University of Wisconsin-Milwaukee, Milwaukee, Wisconsin 53211, USA  (email: andrei.nemilentsau@gmail.com).

George W. Hanson is with Department of Electrical Engineering, University of Wisconsin-Milwaukee, Milwaukee, Wisconsin 53211, USA. (email: george@uwm.edu).

}}

\markboth{Journal of \LaTeX\ Class Files,~Vol.~xxx, No.~xxx, xxx~2015}%
{Shell \MakeLowercase{\textit{et al.}}: Bare Demo of IEEEtran.cls for Journals}



% for Transactions on Magnetics papers, we must declare the abstract and
% index terms PRIOR to the title within the \IEEEtitleabstractindextext
% IEEEtran command as these need to go into the title area created by

\maketitle

\begin{abstract}
We study the electromagnetic response of two- and quasi-two-dimensional hyperbolic materials, on which a simple dipole source can excite a well-confined and tunable surface plasmon polariton (SPP). The analysis is based on the Green's function for an anisotropic two-dimensional surface, which nominally requires the evaluation of a two-dimensional Sommerfeld integral. We show that for the SPP contribution this integral can be evaluated efficiently in a mixed continuous-discrete form as a continuous spectrum contribution (branch cut integral) of a residue term, in distinction to the isotropic case, where the SPP is simply given as a discrete residue term. The regime of strong SPP excitation is discussed, and complex-plane singularities are identified, leading to physical insight into the excited SPP. We also present a stationary phase solution valid for large radial distances. Examples are presented using graphene strips to form a hyperbolic metasurface, and thin-film black phosphorus. The 
 Green's function and complex-plane analysis developed allows for the exploration of hyperbolic plasmons in general 2D materials.
\end{abstract}


\maketitle

\IEEEdisplaynontitleabstractindextext

% \fi
\IEEEpeerreviewmaketitle

\section{Introduction}

\IEEEPARstart{R}{}ecently, the development of nano-fabrication technologies has made it possible to fabricate artifical materials exhibiting a hyperbolic regime - hyperbolic metamaterials (HMTMs) \cite{IEEEhowto:Podolskiy,IEEEhowto:Capolino}. Hyperbolic metamaterials are uniaxial structures with extreme anisotropy, whose reactive effective material tensor components have opposite signs for orthogonal electric field polarizations \cite{IEEEhowto:Alu}. Hyperbolic materials exhibit hyperbolic, as opposed to the usual elliptic, dispersion, and combine the properties of transparent dielectrics and reflective metals \cite{IEEEhowto:Podolskiy}. These exotic properties have led to new physical phenomena and to the proposal for optical devices for a wide range of applications, such as far-field subwavelength imaging, nanolithography, emission engineering \cite{IEEEhowto:Podolskiy}, negative index waveguides \cite{IEEEhowto:Narimanov}, subdiffraction photonic funnels \cite{IEEEhowto:Govyadinov}, and nanoscale resonators \cite{IEEEhowto:Bartal}. 

The complexity of bulk fabrication of metamaterials has hindered the impact of this technology, especially in the optical regime, and volumetric effects may be detrimental for the associated losses \cite{IEEEhowto:Alu}. Metasurfaces \cite{Kuester}-\cite{IEEEhowto:Capasso}, sheets of material with extreme sub-wavelength thickness, might address many of the present challenges and allow integration with planarized systems compatible with integrated circuits. Many high frequency electronics applications are envisioned for metasurfaces due to their ability to support and guide highly confined surface plasmons. The class of two dimensional (2D) atomic crystals \cite{IEEEhowto:Novoselov} represents the ultimate embodiment of a meta-surface in terms of thinness, and often performance (e.g. tunability, flexibility, quality factor). Some notable examples of 2D layered crystals include graphene, transition metal dichalcogenides, trichalcogenides, black phosphorus, boron nitride, and many more. 

Graphene in particular has received considerable attention as a promising two-dimensional surface for many applications relating to large enhancement in Purcell emission, integrability, electronic tunability and tranformation optics \cite{Tony1}-\cite{IEEEhowto:Engheta}. In addition to graphene, black phosphorus (BP) is also a layered material, with each layer forming a puckered surface due to sp3 hybridization. It is one of the thermodynamically more stable phases of phosphorus, at ambient temperature and pressure \cite{IEEEhowto:Morita}. BP has recently been exfoliated into its multilayers \cite{IEEEhowto:L.Li}-\cite{IEEEhowto:Koenig}, showing good electrical transport properties. In particular, the optical absorption spectra of BP vary sensitively with thickness, doping, and light polarization, especially across the technologically relevant mid- to near-infrared spectrum \cite{IEEEhowto:Tony3}-\cite{IEEEhowto:Qiao}. Hence, it has also received considerable attention for optoelectronics, such as hyperspectral imaging and detection \cite{IEEEhowto:Tony2}-\cite{IEEEhowto:Yuan}, photodetectors in silicon photonics \cite{IEEEhowto:Youngblood}, photo-luminescence due to excitonic effects \cite{IEEEhowto:Castellanos}, among many others. 

Both natural  materials and metasurfaces can be isotropic or anisotropic, and, e.g., isotropic graphene can be employed to form an effective anisotropic metasurface by modulating its conductivity \cite{IEEEhowto:Alu, IEEEhowto:Forati}. And, both natural materials and metasurfaces may exhibit a hyperbolic regime. Basic properties of plasmons on 2D hyperbolic surfaces have been recently studied; for metasurfaces comprised of anisotropic plasmonic particles in \cite{Yer}, for graphene strips in \cite{IEEEhowto:Alu}, and for general continuum 2D materials including black phosphorus in \cite{AN}.

In this work we provide the Green's function for an anisotropic two-dimensional surface in Sommerfeld integral form. We focus on complex-plane analysis of the Green's function for the SPP contribution in the hyperbolic case. The nominally two-dimensional Sommerfeld integral form of the Green's function is very time-consuming to evaluate, and provides no physical insight into the resulting field. Here we show that for the SPP field, this integral can be evaluated efficiently in a mixed continuous-discrete form as a continuous spectrum contribution (branch cut integral) of a residue term. Complex-plane singularities are identified with various branch cut integrals, leading to physical insight into the excited SPP. For some two-dimensional materials the surface conductivity is rather weak, and a discussion is provided concerning the strength of the reactive conductivity response to maintain an SPP. 

The paper is organized as follows. We discuss the Green's function calculation for an anisotropic two-dimensional sheet with conductivity tensor $ \underline{\sigma} $. A Hertzian dipole vertical current source serves as the excitation. Rigorous complex plane analysis is shown to reduce the two-dimensional iterated Sommerfeld integral to a residue for the inner integral (for the SPP contribution), and a branch cut for the outer integral \cite{IEEEhowto:Chang}-\cite{IEEEhowto:Dennis}. The relevant singularities are detailed. We also provide a stationary phase (SP) evaluation leading to a closed-form solution for large radial distances. We show that graphene strips support propagation of directed surface waves and that the direction of propagation can be controlled by changing the frequency or doping. We also consider black phosphorus, which is dynamically tunable and anisotropic, and can be hyperbolic.

\section{Fundamental Equations}
\begin{figure}[h]
	\begin{center}
		\noindent
		\includegraphics[width=2in]{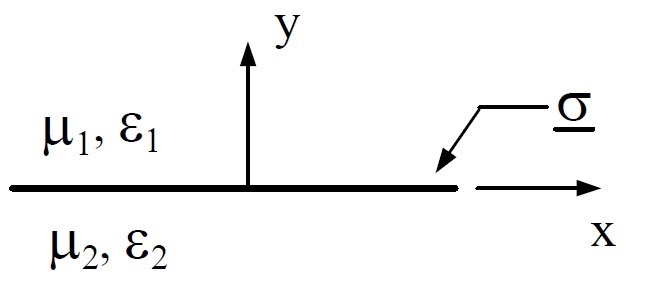}
		\caption{Anisotropic  surface with conductivity tensor $ \overline{\sigma} $ at the interface of two isotropic materials.}\label{Fig1}
	\end{center}
\end{figure}

The geometry under consideration is shown in Fig. \ref{Fig1}. We consider an anisotropic  layer with conductivity tensor
\begin{equation}\label{Eq:1}
 \underline{\sigma}=	\left( 
 \begin{array}{ccc}
 \sigma_{xx}& 0 \\ 
 0 & \sigma_{zz}%
 \end{array}%
 \right)
\end{equation} 
embedded at the interface of two isotropic different materials with electrical properties $ \epsilon_1 $, $ \mu_1 $ and $ \epsilon_2 $, $ \mu_2 $.

%\subsection{Green's Function}

For any planarly layered, piecewise-constant medium, the electric and
magnetic fields in region $n$ due to an electric current in any region are 
\begin{align}
\mathbf{E}^{\left( n\right) }\left( \mathbf{r}\right) & =\left( k_{n}^{2}+\mathbf{\nabla \nabla \cdot }\right) {\boldsymbol \pi }^{\left( n\right) }\left( 
\mathbf{r}\right)   \label{ffp} \\
\mathbf{H}^{\left( n\right) }\left( \mathbf{r}\right) & =i\omega \varepsilon_{n}\mathbf{\nabla }\times {\boldsymbol \pi }^{\left( n\right) }\left( \mathbf{r}
\right)   \label{fpp1}
\end{align}
where $k_{n}=\omega \sqrt{\mu _{n}\varepsilon _{n}}$ and $\mathbf{\pi }%
^{\left( n\right) }\left( \mathbf{r}\right) $ are the wavenumber and
electric Hertzian potential in region $n$, respectively. The suppressed time convention is $ e^{i\omega t} $. Assuming that the
current source is in region 1, $ \mathbf{J}^{\left(1\right)} $, then%
\begin{align}
{\boldsymbol \pi }^{\left( 1\right) }\left( \mathbf{r}\right) & ={\boldsymbol \pi }_{1}^{p}\left( \mathbf{r}\right) +{\boldsymbol \pi }_{1}^{s}\left( \mathbf{r}\right) \notag \\
& =\int_{\Omega }\left\{ \underline{\mathbf{g}}^{p}\left( \mathbf{r,r}^{\prime }\right) +\underline{\mathbf{g}}^{r}\left( \mathbf{r,r}^{\prime}\right) \right\} \cdot \frac{\mathbf{J}^{\left( 1\right) }\left( \mathbf{r}^{\prime }\right) }{i\omega \varepsilon _{1}}\,d\Omega ^{\prime } \notag \\ 
{\boldsymbol \pi }^{\left( 2\right) }\left( \mathbf{r}\right) & ={\boldsymbol \pi }_{2}^{s}\left( \mathbf{r}\right) =\int_{\Omega }\underline{\mathbf{g}}^{t}\left( \mathbf{r,r}^{\prime }\right) \cdot \frac{\mathbf{J}^{\left(1\right) }\left( \mathbf{r}^{\prime }\right) }{i\omega \varepsilon _{1}}\,d\Omega ^{\prime }  \label{hpt}
\end{align}\label{Eq:4}
where the underscore indicates a dyadic quantities, $ \underline{\mathbf{g}}^{p} $ is the principal (free space) dyadic Green's function, $ \underline{\mathbf{g}}^{r} $ is the reflected dyadic Green's function responsible for the fields in the region containing the source, $ \underline{\mathbf{g}}^{t} $ is the transmitted dyadic Green's function responsible for the fields in the non-source region (here we assume a source in one region or the other, but not in both regions) and $\Omega $ is the
support of the current. With $y$ parallel to the interface normal, the
principle Green's dyadic can be written as 
\begin{align}
&\underline{\mathbf{g}}^{p}\left( \mathbf{r,r}^{\prime }\right) =\underline{%
\mathbf{I}}\,\frac{e^{-ik_{1}R}}{4\pi R}  \notag\\& 
~~~~~~~~~~~=\underline{\mathbf{I}}\,\frac{1}{\left( 2\pi \right) ^{2}}\int_{-\infty
}^{\infty }\int_{-\infty }^{\infty }\frac{e^{-p_{1}\left\vert y-y^{\prime
}\right\vert }}{2p_{1}}e^{-i\mathbf{q}\cdot \left( \mathbf{r-r}^{\prime
}\right) }\,dq_{x}dq_{y}  \label{gp4}
\end{align}\label{Eq:5}
where $ \mathbf{q}=\widehat{\mathbf{x}}q_{x}+\widehat{\mathbf{z}}q_{z} $, $\left\vert \mathbf{q}\right\vert =q=\sqrt{q_{x}^{2}+q_{z}^{2}}$, $ p_{n}^{2}=\left\vert \mathbf{q}\right\vert^{2}-k_{n}^{2} $, $ \rho =\sqrt{\left( x-x^{\prime
	}\right) ^{2}+\left( z-z^{\prime }\right) ^{2}} $, $ R=\left\vert \mathbf{r-r}^{\prime }\right\vert =\sqrt{\rho ^{2}+\left(
	y-y^{\prime }\right) ^{2}} $ and $\underline{\mathbf{I}}$ is the unit dyadic.

The scattered (reflected or transmitted) Green's dyadics can be obtained by enforcing the boundary
conditions 
\begin{align}
	\widehat{\mathbf{z}}\times \left( \mathbf{H}_{1}-\mathbf{H}_{2}\right) & =%
	\mathbf{J}_{e}^{s}  \notag \\
	\widehat{\mathbf{z}}\times \left( \mathbf{E}_{1}-\mathbf{E}_{2}\right) & =-%
	\mathbf{J}_{m}^{s}  \label{BCME}
\end{align}\label{Eq:6}
\noindent where $\mathbf{J}_{e}^{s}$ (A/m) and $\mathbf{J}_{m}^{s}$ (V/m) are electric
and magnetic surface currents on the boundary. In our case, $\mathbf{J}_{m}^{s}=\mathbf{0}$, and $\mathbf{J}_{e}^{s}=\underline{\mathbf{\sigma }}\cdot \mathbf{E}$. Using only an electric Hertzian potential, we can satisfy Maxwell's equations and the relevant boundary conditions. Introducing the two-dimensional Fourier transform 
\begin{align}\label{Eq:7}
	\mathbf{a}\left( \mathbf{q},y\right) & =\int_{-\infty }^{\infty
	}\int_{-\infty }^{\infty }\mathbf{a}\left( \mathbf{r}\right) e^{i\mathbf{%
		q\cdot r}}\,dxdz \\
\mathbf{a}\left( \mathbf{r}\right) & =\frac{1}{\left( 2\pi \right) ^{2}}%
\int_{-\infty }^{\infty }\int_{-\infty }^{\infty }\mathbf{a}\left( \mathbf{q}%
,y\right) e^{-i\mathbf{q\cdot r}}\,dq_{x}dq_{z}
\end{align}
and enforcing the boundary conditions, the scattered Green's dyadic is found to
have the form%
\begin{equation}\label{Eq:9}
	\underline{\mathbf{g}}^{r,t}=\left( 
	\begin{array}{ccc}
		g_{xx}^{r,t} & g_{xy}^{r,t} & 0 \\ 
		g_{yx}^{r,t} & g_{yy}^{r,t} & g_{yz}^{r,t} \\ 
		0 & g_{zy}^{r,t} & g_{zz}^{r,t}%
	\end{array}%
	\right)
\end{equation}%
\noindent where the Sommerfeld integrals are%
\begin{align}\label{Eq:10}
&g_{\alpha \beta }^{r}\left( \mathbf{r,r}^{\prime }\right) = \notag \\ & 
\frac{1}{\left(2\pi \right) ^{2}}\int_{-\infty }^{\infty }\int_{-\infty }^{\infty
}w_{\alpha \beta }^{r}\left( q_{x},q_{z}\right) \frac{e^{-p_{1}\left(
y+y^{\prime }\right) }}{2p_{1}}e^{-i\mathbf{q\cdot }\left( \mathbf{r-r}%
^{\prime }\right) }\,dq_{x}dq_{z}
\end{align}
The Green's dyadic for region 2, $\underline{\mathbf{g}}^{t}\left( \mathbf{r,r}^{\prime }\right) $, has the same form as for region 1, although in (\ref{Eq:10}) the replacement $ w_{\alpha \beta }^{r}e^{-p_{1}\left( y+y^{\prime }\right) }\rightarrow
w_{\alpha \beta }^{t}e^{p_{2}y}e^{-p_{1}y^{\prime }} $ must be made.

The coefficients $w_{\alpha \beta}^{r,t}$ are complicated for the inhomogeneous case, and so for simplicity in the following we assume the sheet is in a homogeneous space $\varepsilon _{2}=\varepsilon _{1}=\varepsilon$, $\mu _{2}=\mu _{1}=\mu$. When region 2 differs from region 1, the only change is in the functions (\ref{Eq:13aa})-(\ref{Eq:13}) provided below. Concentrating on the field in the upper-half space, $w_{\alpha \beta}^{r}={N_{\alpha \beta}\left( q_{x},q_{z}\right) }/{D\left( q_{x},q_{z}\right) }, $ where

\begin{align}\label{Eq:13aa}
D\left( q_{x},q_{z}\right)=2\sigma _{xx}&\left( k^{2}-q_{x}^{2}\right)
+2\sigma _{zz}\left( k^{2}-q_{z}^{2}\right) \notag \\& 
-i4\frac{k}{\eta }p\left( 1+\frac{1}{4}\eta ^{2}\sigma _{xx}\sigma
_{zz}\right),
\end{align}
and
\begin{align}\label{Eq:13}
&N_{yy}\left( q_{x},q_{z}\right)= -p^{2}\left( \sigma _{xx}+\sigma _{zz}\right) -ip k\eta
\sigma _{xx}\sigma _{zz}, \notag \\
&N_{xy}\left( q_{x},q_{z}\right)= i q_x p\left( \sigma _{xx}-\sigma _{zz}\right), \notag \\
&N_{zy}\left( q_{x},q_{z}\right)= -i q_z p\left( \sigma _{xx}-\sigma _{zz}\right), \notag \\
\end{align}
\noindent where $p=\sqrt{q_{x}^{2}+q_{z}^{2}-k^{2}}$, and $\eta = \sqrt{\mu/\varepsilon}$. Then, e.g., for the vertical field in the upper half-space,

\begin{align}\label{Eq:14}
	&E_y =\frac{1}{i \omega \epsilon}   \left(k^2+ \frac{\partial^2}{\partial y^2} \right)\left(g_{yy}^p\left( \mathbf{r,r}^{\prime }\right)+g_{yy}^r\left( \mathbf{r,r}^{\prime }\right)\right)  \notag \\&
	~~~~~~+ \frac{1}{i \omega \epsilon} \left( \frac{\partial^2}{\partial x \partial y} g_{xy}^r\left( \mathbf{r,r}^{\prime }\right)  + \frac{\partial^2}{\partial z \partial y} g_{zy}^r\left( \mathbf{r,r}^{\prime } \right)  \right) 
	\end{align}
and other field components are obtained from (\ref{ffp}).

\section{Directional properties of SPPs on 2D surfaces}

\begin{figure}[h]
  \begin{center}
		\noindent
       \includegraphics[width=3.6in]{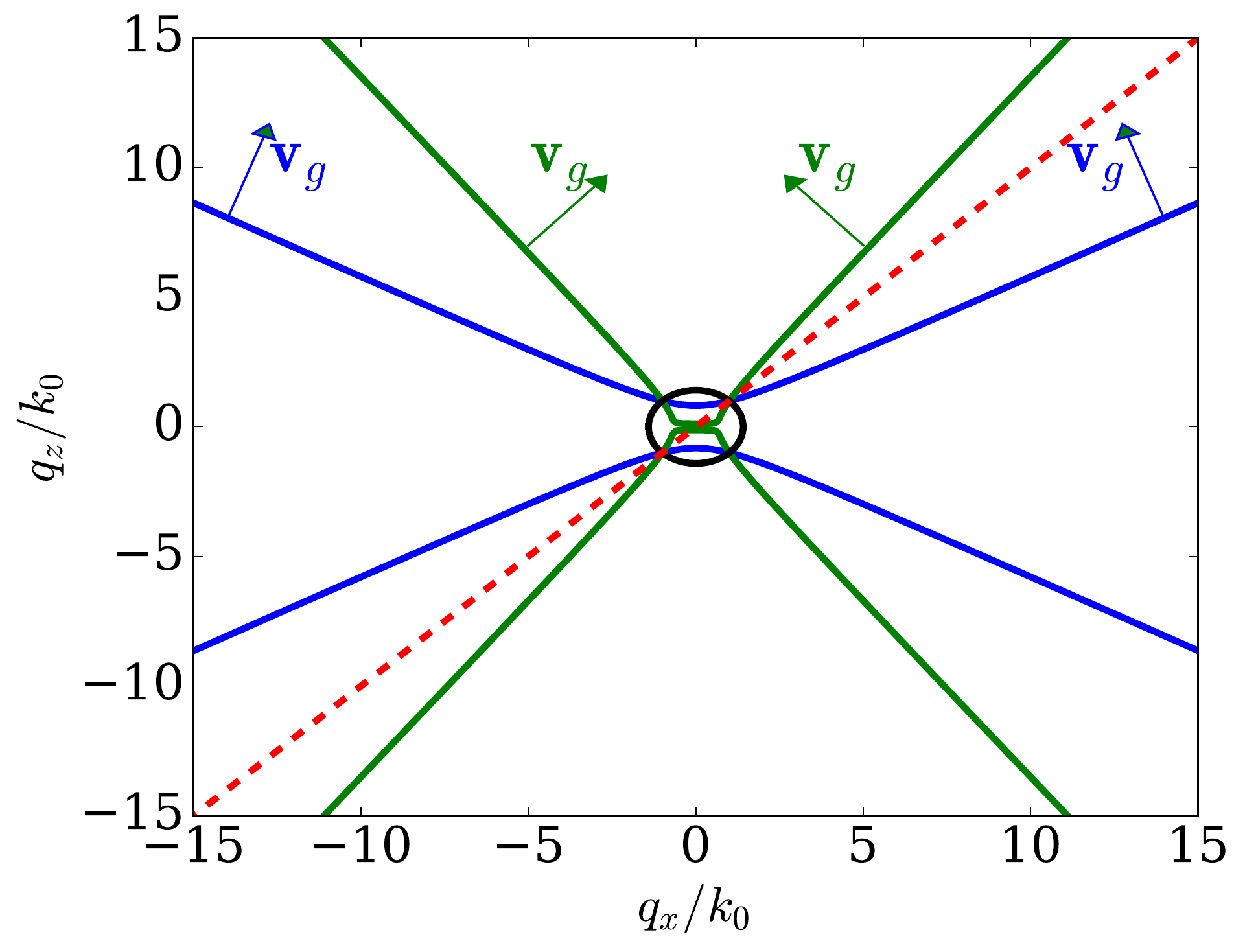}
       \caption{Equifrequency surfaces for metasurface having $\sigma_{xx} =0.003 + 0.25i$ mS and $ \sigma_{zz} =0.03 - 0.76i$ mS (blue hyperbola; see also Fig. \ref{Fig8}b), and $ \sigma_{xx} = 1.3 +16.9i  $ mS and $ \sigma_{zz} = 0.4 - 9.2i  $ mS (green hyperbola; see also Fig. \ref{Fig8}c). For comparison, the isotropic case for $ \sigma_{xx} = \sigma_{zz} =0.03 - 0.76i$ mS (black circle) is also shown. The red dashed line merely denotes 45 degrees with respect to the $x$ axis for guidance. }
\label{equi}
\end{center} 
\end{figure}

Before considering complex-plane evaluation of the Green's functions, we describe some basic properties of SPPs on hyperbolic 2D surfaces \cite{IEEEhowto:Alu}, \cite{Yer}-\cite{AN}. In order to understand the behavior of surface waves it is instructive to inspect the plasmon dispersion relation $D\left( q_{x},q_{z}\right)=0$ arising from (\ref{Eq:13aa}), the denominator of the Green's function. As we show later, in the general case SPPs are obtained as a mixture of TE and TM modes, and, moreover, it is not possible to solve for the wavevector eigenmodes $q_x$ and $q_z$ from the single complex-valued equation (\ref{Eq:13aa}). Furthermore, unlike for isotropic surfaces, for an anisotropic medium the direction of energy transfer is defined by the group velocity in the medium \cite{kong}, $\nabla_{\mathbf{q}} \omega(\mathbf{q})$, and does not coincide with the direction of the plasmon wavevector $\mathbf{q}$. In our case, the dispersion relation for surface plasmons is complicated and the group velocity can not be calculated analytically. However, we can estimate the direction of plasmon propagation geometrically by examining the plasmon's equifrequency contours, $\omega(\mathbf{q}) = const$. As the group velocity is a gradient of frequency with respect to wavevector, the direction of plasmon energy flow is necessary orthogonal to the equifrequency contours.

Assuming that the conductivity is purely imaginary and lossless, $\sigma_{jj} = i\sigma''_{jj}$, $j=x,z$, and that $q_x, q_z \gg k$, the zeros of (\ref{Eq:13aa}) can be approximated as the solution of
\begin{align}
\frac{q_x^2}{\sigma''_{zz}}  + \frac{q_z^2}{\sigma''_{xx}}  = 2 p \omega \left(\frac{\varepsilon_0}{\sigma''_{xx} \sigma''_{zz}} - \frac{\mu_0 }{4} \right). \label{Eq:dispersion1}
\end{align}

Although the right side varies with $\mathbf{q}$, because of the square-root $p$ the variation is less than the left side, and we can approximate the right side as being constant in wavenumber. Then, in the hyperbolic case ($\sigma''_{xx}\cdot\sigma''_{zz} < 0$) the EFS is a hyperbola, as shown in Fig. \ref{equi} for two values of surface conductivity (blue lines: $\sigma_{xx} =0.003 + 0.25i$ mS and $ \sigma_{zz} =0.03 - 0.76i$ mS; see also Fig. \ref{Fig8}b, and green lines: $ \sigma_{xx} = 1.3 +16.9i  $ mS and $ \sigma_{zz} = 0.4 - 9.2i  $ mS; see also Fig. \ref{Fig8}c); results in Fig. \ref{equi} were obtained by solution of the full dispersion relation (\ref{Eq:13aa}). The hyperbola asymptotes are defined by $q_z = \pm q_x \sqrt{|\sigma''_{xx}/\sigma''_{zz}|}$. Taking into account that a dipole excites many plasmons with different $\mathbf{q}$, and that the normal to all the points on the hyperbola point in the same direction for a given sign of $q_x$, we expect a narrow plasmon beam in the direction of energy flow on a hyperbolic metasurface. For example, the asymptotes of the blue hyperbola in Fig. \ref{equi} have an angle 30 degrees with respect to the $x$ axis, and thus the normal to the hyperbola, i.e., the group velocity, is 60 degrees with respect to the $x$ axis, as indicated in the figure, which is in very good agreement with the numerical results presented in Fig. \ref{Fig8}b. Similar comments apply to the green hyperbola and Fig. \ref{Fig8}c. For comparison, in Fig. \ref{equi} we also presented the hypothetical isotropic case for which the equifrequency contour is a circle, and thus energy does not have a preferential direction.   

In the non-hyperbolic (purely anisotropic) case ($\sigma''_{xx}, \sigma''_{zz} > 0$), \eqref{Eq:dispersion1} is the equation for an ellipse in $\mathbf{q}$-space with the axis oriented along $q_x$ and $q_z$. The length of the ellipse's principal axes along $q_x$ and $q_z$ is proportional to $\sigma''_{zz}$ and $\sigma''_{xx}$, respectively. Thus, the EFS has a quasi-eliptic form elongated along the direction of the smallest component of the conductivity tensor, the degree of elongation being set by the ratio of $\sigma''_{xx}$ and $\sigma''_{zz}$. Later, in Fig. \ref{Fig9} we consider black phosphorous having $ \sigma_{xx} = 0.0008 - 0.2923i  $ mS and $ \sigma_{zz} =  0.0002 - 0.0658i $ mS. Due to the strong elongation of the EFS along the $q_z$-axis, the group velocity points approximately along the $q_x$ axis, such that the SPP carries energy along the $x$ crystallographic axis (see, e.g., Fig. \ref{Fig9}). 

\section{Complex-Plane Analysis in the $q_x$-Plane}

In the case of an isotropic material the coefficients $w_{\alpha \beta}$ only depend
on $q^{2}=q_{x}^{2}+q_{z}^{2}$, leading
to  
\begin{align}\label{Eq:15}
&g_{\alpha \beta}^{r}\left( \mathbf{r,r}^{\prime }\right)  =\frac{1}{2\pi }%
\int_{0}^{\infty }w_{\alpha \beta}\left( q\right) \frac{e^{-p\left( y+y^{\prime
}\right) }}{2p}J_{0}\left( q\rho \right) qdq  \notag \\&
~~~~~~~~~~~~ =\frac{1}{2\pi }\int_{-\infty }^{\infty }w_{\alpha \beta}\left( q\right) \frac{e^{-p\left( y+y^{\prime }\right) }}{4p}H_{0}^{\left( 2\right) }\left( q\rho
\right) qdq 
\end{align}%
where $J_0$ and $H_0^{(2)}$ are the usual zeroth-order Bessel and Hankel functions, respectively. These two forms can be converted one to another using the relation $ J_0(\alpha) = \frac{1}{2} \left[ H_0^{(1)}(\alpha) + H_0^{(2)}(\alpha)  \right], ~ H_0^{(2)}(-\alpha) = -H_0^{(1)}(\alpha)  $. In this case, such as occurs for graphene without a magnetic bias, the pole of $w_{\alpha \beta}$ leads to a simple analytical form for the SPP field \cite{IEEEhowto:Hanson}. However, this is not the case for an anisotropic surface. Since the two-dimensional Sommerfeld integral can be time-consuming to evaluate, writing

\begin{equation}\label{Eq:16}
g_{\alpha \beta}^{r}\left( \mathbf{r,r}^{\prime }\right) =\frac{1}{\left( 2\pi \right) 
}\int_{-\infty }^{\infty }dq_{z}e^{-iq_{z}\left( z-z^{\prime }\right)
}f_{\alpha\beta}\left( q_{z}\right) 
\end{equation}%
where 
\begin{equation}\label{Eq:17}
f_{\alpha \beta}\left( q_{z}\right) =\frac{1}{\left( 2\pi \right) }\int_{-\infty }^{\infty }%
w_{\alpha \beta}(q_x,q_z)\frac{%
e^{-p\left( y+y^{\prime }\right) }}{2p}e^{-iq_{x}\left( x-x^{\prime }\right)
}dq_{x}
\end{equation}%
the ``inner" integral $f_{\alpha \beta}\left( q_{z}\right) $ can be evaluated as an SPP residue term (discrete spectral component) and branch cut integral representing the radiation continuum into space (note that the choice of ``inner" and ``outer" integrals is arbitrary). The branch cut in the $q_x$ plane is the usual hyperbolic branch cut associated with the branch point due to $p=\sqrt{q_{x}^{2}+q_{z}^{2}-k^{2}}$, occurring at $q_{x}=\pm \sqrt{k^{2} - q_{z}^{2}}$ \cite{Ishimaru}. Then,

\begin{align}\label{Eq:18}
	&f_{\alpha \beta}\left( q_{z}\right)=-i w^{spp}_{\alpha \beta}(q_{xp},q_z) \frac{%
		e^{-p\left( q_{xp}\right) \left( y+y^{\prime }\right) }}{2p\left(
		q_{xp}\right) }e^{-iq_{xp}\left( x-x^{\prime }\right) } \notag \\&
	~~~~~~~~~~+\frac{1}{2\pi}\int_{\text{bc}}
	w_{\alpha \beta}(q_x,q_z)\frac{%
		e^{-p\left( y+y^{\prime }\right) }}{2p}e^{-iq_{x}\left( x-x^{\prime
		}\right) }dq_{x} 
	\end{align}
where the first term is the residue contribution and bc indicates the hyperbolic branch-cut contour. In (\ref{Eq:18}), $w^{spp}(q_{xp},q_z)=N(q_{xp},q_z)/D'(q_{xp},q_z)$, $D'(q_x,q_z)=\frac{\partial}{\partial q_{x}}D\left( q_{x},q_{z}\right) $, and where $ q_{xp} $ is the root of $ D(q_{x},q_z)=0 $ for a given $ q_z $,

\begin{equation}\label{Eq:19}
	q_{xp}(q_z)=\pm\sqrt{\frac{-B \pm \sqrt{B^2-4AC}}{2A}}
\end{equation}

\noindent where $A=\sigma^2_{xx}$, $B=\frac{1}{4} \alpha^2 -2k^2\sigma^2_{xx}+2(q^2_z-k^2)\sigma_{xx}\sigma_{zz}$, $C=k^4(\sigma_{xx}+\sigma_{zz})^2+q^2_z(q^2_z-2k^2)\sigma^2_{zz}-2k^2q^2_z\sigma_{xx}\sigma_{zz} +\frac{1}{4}\alpha^2(q^2_z-k^2)$, and $\alpha=\left (4k/ \eta \right ) (1+\frac{1}{4}\eta^2\sigma_{xx}\sigma_{zz})$. 

When the SPP field is the dominant contribution to the response, which is the usual regime for plasmonics where the field close to the interface, ($y,y' \ll \lambda$)) is of interest, the branch cut term can be ignored and the residue term suffices for the calculation of $f(q_z)$,
\begin{equation}\label{Eq:20a}
f_{\alpha \beta}^{\text{SPP}}\left( q_{z}\right) \approx -i w^{spp}_{\alpha \beta}(q_{xp},q_z) \frac{%
e^{-p\left( q_{xp}\right) \left( y+y^{\prime }\right) }}{2p\left(
q_{xp}\right) }e^{-iq_{xp}\left( x-x^{\prime }\right) },
\end{equation}
which considerably speeds up evaluation of the Green's function (rendering it one-dimensional). Since $q_{xp}$ is the propagation constant along the $x$-axis, the $ \mp $ outside the square root in (\ref{Eq:19}) indicates forward/backward propagation, whereas the inner $ \pm $ sign choice governs propagation of different modes (only one of which will propagate). Assuming $(x - x') > 0$, the term $e^{-iq_{xp}(x-x')}$ necessitates that $\mathrm{Im}(q_{xp})<0$ to have a decaying wave traveling away from the source along the $x$-axis.  

\begin{figure}[h]
	\begin{center}
		\noindent
		\includegraphics[width=3.6in]{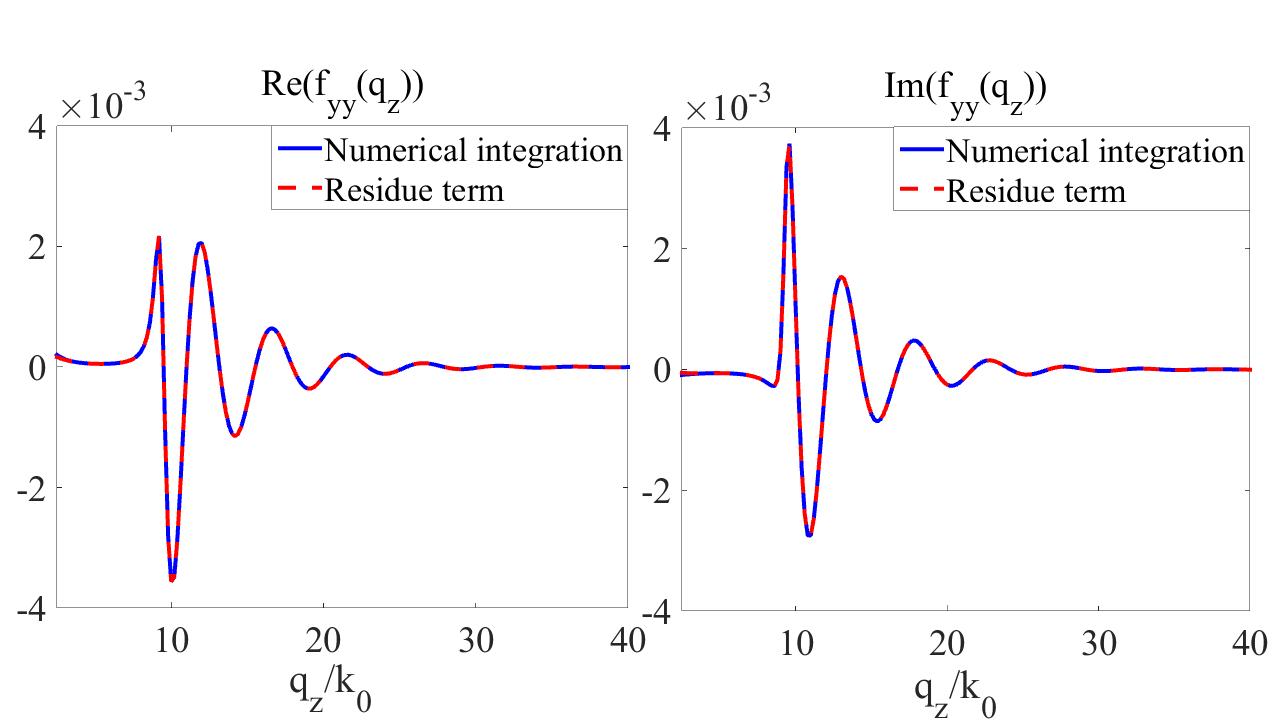}
		\caption{Real and imaginary parts of $ f_{yy}(q_z) $ obtained numerically, (\ref{Eq:17}), and using the residue term (\ref{Eq:20a}) for an array of graphene strips at $ f=10 $ THz. The source is $ \lambda/50 $ above the surface, and $ x= 0.2 \lambda $.}\label{Fig2}
	\end{center}
\end{figure}

As an example, we consider an anisotropic surface with $ \sigma_{xx}=0.02 + 0.57i $ mS and $ \sigma_{zz} = 0.02 - 0.57i $ mS. As discussed in Appendix A, such a conductivity tensor can be physically realized by an array of densely packed graphene strips at terahertz and near infrared frequencies. Fig.~\ref{Fig2} compares $ f_{yy} (q_z) $ obtained numerically by performing the integral (\ref{Eq:16}) and obtained by using the residue term only, (\ref{Eq:20a}). The source is located at $y'=\lambda/50 $, very near the surface, and radiating at frequency $ 10 $ THz. Clearly, in the SPP regime the residue provides the dominat component of the response, and the branch cut integral can be ignored. Although not shown, for source or observation points relatively far from the surface, the branch cut integral is important, and can be the dominant contribution to the scattered field. 

\begin{figure}[h]
	\begin{center}
		\noindent
		\includegraphics[width=3.1in]{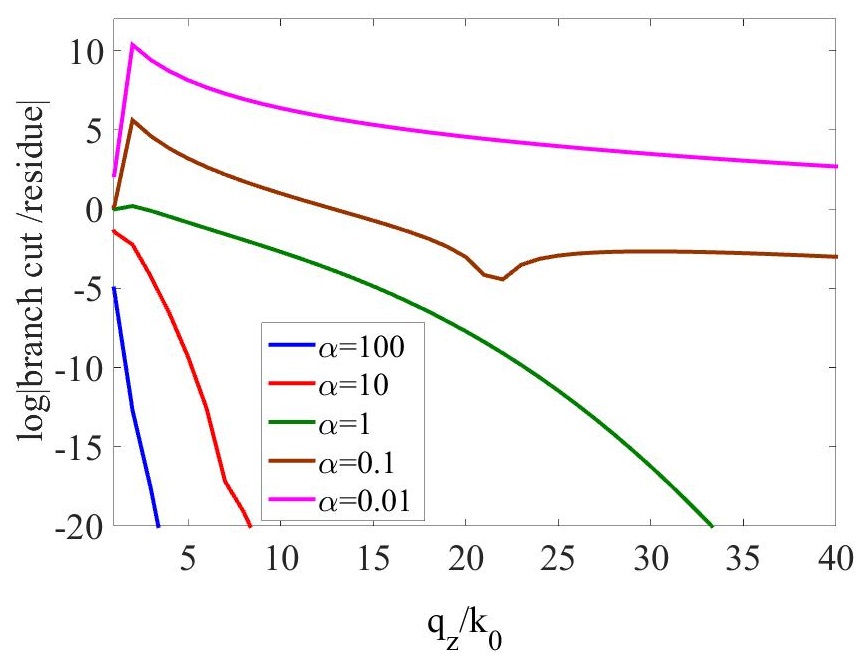}
		\caption{Ratio of the branch cut and residue terms in (\ref{Eq:18}), $ \sigma_{xx}= \alpha\sigma_0(0.01+i), \sigma_{zz}=0.1 \sigma^*_{xx} , \sigma_0=e^2/4\hbar$. Source is positioned $ \lambda/50 $ above the surface, $ f=10 $ THz, and $ x= 0.2 \lambda $. }\label{Fig6}
	\end{center}
\end{figure}

In the following we are interested in surfaces that provide a strong reactive and low-loss response, $\text{Im}(\sigma_{\alpha \alpha})\gg \text{Re}(\sigma_{\alpha \alpha})$. In addition to this inequality, $\text{Im}(\sigma_{\alpha \alpha})$ must not be too small \cite{Hanson2015}. The ability of a surface to support a strong SPP depends on the ratio of the branch cut term (space radiation spectra) to the residue (SPP) term in the inner integral (\ref{Eq:18}). In Fig. \ref{Fig6} we assume a general hyperbolic form $ \sigma_{xx}=\alpha \sigma_0(0.01+i) $ and $ \sigma_{zz}=0.1 \sigma_{xx}^* $ where $ \sigma_0=e^2/4 \hbar $ is the conductance quantum, $ e $ is the electron charge, and $^*$ indicates complex conjugation. We assume that losses are relatively small, and use $\alpha$ in order to vary the magnitude of the conductivity.

It is clearly shown in Fig. \ref{Fig6} that for conductivity values smaller than the conductance quantum, the radiation spectra is dominant (in the limit that $|\sigma|\rightarrow 0$, the surface vanishes and the entire response is the radiation continuum produced by a source in free space). We have found that conductivity values on the order of the conductance quantum are somewhat borderline; an SPP can exist, although it may not be strongly dominant over the branch cut continuum for small $q_z$. Conductivities an order of magnitude or more above the conductance quantum provide a very strong SPP response in which the branch cut contribution is negligible except exceedingly close to  the source. 

For large $ q_z $ compared to $k$, (\ref{Eq:19}) becomes 

\begin{equation}\label{Eq:21}
q_{xp}(q_z)=q_z \sqrt{-\frac{\sigma_{zz}}{\sigma_{xx}}}.
\end{equation}
The SPP direction of propagation on the 2D anisotropic surface is easily determined as $ \tan^{-1}(\frac{q_{xp}}{q_z}) $, and using (\ref{Eq:21}) the angle of propagation with respect to the $z$-axis is simply

\begin{equation}\label{Eq:22}
\phi=\tan^{-1} \sqrt{-\frac{\sigma''_{zz}}{\sigma''_{xx}}},
\end{equation}
where $\sigma''=\mathrm{Im}({\sigma})$. Although the conductivities are complex-valued, for the low-loss cases of interest we can estimate the real angle $\phi$ by only considering their imaginary parts. Therefore, in the anisotropic hyperbolic case the SPP is directed along a specific angle. For the isotropic case $(\sigma_{xx}=\sigma_{zz})$ this does not occur (and (\ref{Eq:22}) does not apply), since in this case $ q_{xp}^2 + q_z^2=q_p^2 $, where $ q_p $ is the radial in-plane wavenumber. If we measure the angle $ \phi $ relative to the positive $ z $-axis, then at each point in the plane of the surface we have $ x=\rho \sin \phi $, $ z=\rho \cos \phi $, $ q_{xp}=q_p \sin \phi  $ and $ q_{z}=q_p \cos \phi  $. For a source at the origin,

\begin{align}\label{Eq:23}
	&e^{-i\mathbf{q\cdot }\left( \mathbf{r-r}^{\prime }\right) }=e^{-i(q_{xp}x+q_zz)}=e^{-iq_p \rho( \cos^2 \phi+ \sin^2 \phi)} =e^{-iq_p \rho}
	\end{align}
which $ e^{-iq_p \rho} $ describes a SPP wave that is radially propagating along all directions in the plane of the surface. However, in the anisotropic case for large $ q_z $,

\begin{align}\label{Eq:24}
	&e^{-i\mathbf{q\cdot }\left( \mathbf{r-r}^{\prime }\right) }=e^{-i(q_{xp}x+q_zz)}=e^{-i(q_z \sqrt{- \frac{\sigma_{zz}}{\sigma_{xx}} } x+q_zz)} \notag \\&
	~~~~~~~~~~~~~=e^{-iq_z \rho( \sqrt{- \frac{\sigma_{zz}}{\sigma_{xx}} } \sin \phi +\cos \phi)}
	\end{align}
\noindent and the maximum of $ ( \sqrt{- \frac{\sigma_{zz}}{\sigma_{xx}} } \sin \phi +\cos \phi) $ determines the angle at which the SPP is directed. It can be simply shown that this angle is (\ref{Eq:22}). This leads to the conclusion that hyperbolic anisotropy, in contrast to the isotropic case, results in a directed SPP, as expected.

As a function of $ \underline{\sigma} $, there are different dispersion scenarios for SPP propagation. The usual elliptic case is obtained when both imaginary parts of the conductivity have the same sign (inductive when $ \mathrm{Im}(\sigma_{xx,zz})<0 $, capacitive otherwise). A graphene sheet with dominant intra-band conductivity term with $ \mathrm{Im}(\sigma_{xx})=\mathrm{Im}(\sigma_{zz})<0 $ is a natural example of an elliptic isotropic sheet that can support a TM omni-directional SPP. The hyperbolic case occurs when the sign of the imaginary parts of the conductivity components are different. As discussed in Appendices A and B, both a graphene strip metasurface (potentially, metal strips as well) and natural black phosphorus can provide a hyperbolic 2D surface. In this case, as shown in (\ref{Eq:22}) and (\ref{Eq:24}), energy propagation is focused along specific directions governed by the conductivity components \cite{IEEEhowto:Alu}.

\section{Approximation of the Outer Integral using Stationary Phase, and Exact Evaluation using the Continuous Spectrum}
Although the SPP field can be evaluated from a numerical 1D integral, (\ref{Eq:16}) with (\ref{Eq:20a}), it is useful to consider other methods of evaluation that are more computationally rapid, and which lead to physical insight into the problem.

\subsection{Stationary Phase Evaluation of the Outer Integral}

The ``outer" integral (\ref{Eq:16}) using (\ref{Eq:20a}) can be approximated by the well-known method of stationary phase \cite{Felsen}. In particular, an analysis similar to that needed here was performed in \cite{BH1969}, where the inner integral is approximated as a residue (ignoring the branch cut contribution, as we do here), and the outer integral is evaluated using SP. Regarding computation of the outer integral, although it seems difficult to show analytically because of the complicated expression (\ref{Eq:19}) for the pole $q_{xp}(q_z)$, numerical tests show that $\mathrm{Re}(q^2_{xp}+q^2_z-k^2)>0$ for small values of $q_z$. Therefore, no leaky waves are encountered for typical parameter values.

Stationary phase evaluation of (\ref{Eq:16}) with (\ref{Eq:20a}), assuming $\rho \gg \left( y+y^{\prime }\right) $, results in, to first order, 
\begin{equation}
g_{\alpha \beta }^{r}\left( \mathbf{r,r}^{\prime }\right) \simeq \sqrt{\frac{%
e^{-i\frac{\pi }{2}}}{2\pi \gamma ^{\prime \prime }\left( q_{s}\right) }}%
w_{\alpha \beta }^{spp}(q_{s})\frac{e^{-p\left( q_{s}\right) \left(
y+y^{\prime }\right) }}{2p\left( q_{s}\right) }\,e^{-i\gamma \left(
q_{s}\right) }\label{sp}
\end{equation}%
where $w_{\alpha \beta }^{spp}(q_{s})=w_{\alpha \beta }^{spp}(q_{xp}\left(
q_{s}\right) ,q_{s})$, $p\left( q_{s}\right) =p\left( q_{xp}\left(
q_{s}\right) ,q_{s}\right) $, and $\gamma \left( q_{z}\right) =-\left(
q_{xp}\left( q_{z}\right) \left( x-x^{\prime }\right) +q_{z}\left(
z-z^{\prime }\right) \right) $, where $q_{s}$ is the root of $d\gamma
/dq_{z}=0$, which can be obtained as the root of a fourth-order polynomial,
or via numerical root search. See \cite{BH1969} for a ray-optical interpretation of the SP result in anisotropic media.

\begin{figure}[h]
	\begin{center}
		\noindent
		\includegraphics[width=3.5in]{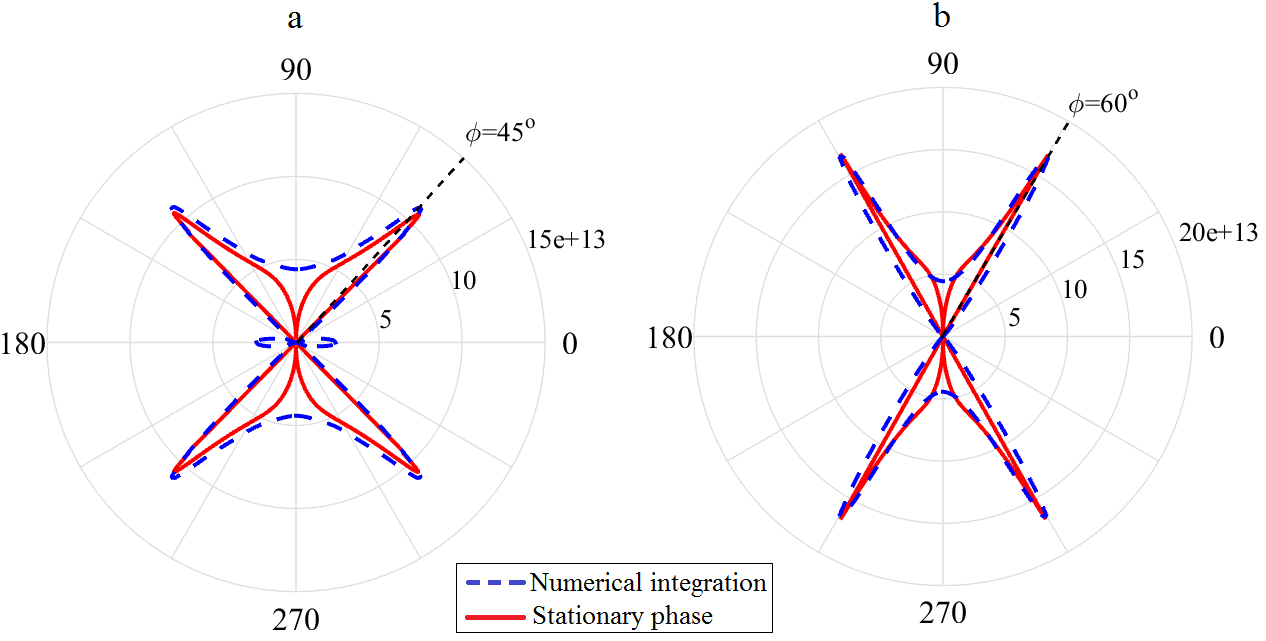}
		\caption{The electric field $ \mathrm{E}_y $ obtained by stationary phase result (\ref{sp}) (red) and numerical integration (\ref{Eq:16})  (blue) for (a) $\sigma_{xx}= 0.02+0.57i \,\text{mS} , \sigma_{zz} = 0.02 - 0.57i \,\text{mS}$ and (b) $\sigma_{xx}= 0.003+0.25i \,\text{mS} , \sigma_{zz} = 0.03 - 0.76i \,\text{mS}$, $\rho=0.4 \lambda$, $\rho/(y+y')=80$, $f=10$ THz. }\label{FigSP}
	\end{center}
\end{figure}

Although the main numerical results will be presented in Section \ref{MR}, here we provide a comparison between the SP result (\ref{sp}) and numerical (real-line) computation of the outer integral (\ref{Eq:16}). Figure \ref{FigSP} shows the SP result (red) and numerical integration result (blue) for (a) $\sigma_{xx}= 0.02+0.57i \,\text{mS} , \sigma_{zz} = 0.02 - 0.57i \,\text{mS}$ and (b) $\sigma_{xx}= 0.003+0.25i \,\text{mS} , \sigma_{zz} = 0.03 - 0.76i \,\text{mS}$, both using $\rho=0.4 \lambda$, $\rho/(y+y')=80$. It can be seen that excellent agreement is found for the location of the beam angle, although away from the beam maximum there is some disagreement.

\subsection{Complex-Plane Analysis in the $q_z-\mathrm{Plane}$}

Although the SPP field can be evaluated to first order using the SP approximation for  $\rho/(y+y') \gg 1$, it is useful to consider complex-plane analysis of the ``outer" integral over $q_z$, which turns out to involve only continuous spectrum. This method is theoretically exact, and is valid for all field and source points. Further, it does not require finding the $q_z$ root, but does require knowing the $q_z$-plane branch points and cuts, which, themselves, lead to considerable physical insight. 

The Weierstrass preparation theorem shows that the complex function $ f^{\text{SPP}}_{\alpha \beta}({q_z}) $, (\ref{Eq:20a}), has no poles, only branch points. Regarding the two complex planes $q_x-q_z$, a sufficient condition in order to have a branch point in the $ q_z-\mathrm{plane} $ is that \cite{For}, \cite{1998}

\begin{equation}
	D\left( q_{x},q_{z}\right) =\frac{\partial}{\partial q_{x}}D\left( q_{x},q_{z}\right) =0
	\label{BP1}
	\end{equation}
with $\delta =\frac{\partial}{\partial q_{z}}D\left( q_{x},q_{z}\right) \frac{\partial^{2}}{\partial q_{x}^{2}}D\left( q_{x},q_{z}\right) \neq 0$. Although (\ref{BP1}) represents a second-order zero of $D$, in the $q_z$-plane these points are not poles, and are also not necessarily $q_z$-plane branch points without the condition $\delta \neq 0$. These branch points are associated with modes in the $q_x$-plane merging at a certain value of $q_z$, forming a second-order zero of $D$. Thus, the pair $(q_x,q_z)$ satisfying (\ref{BP1}) and $\delta \neq 0$ represent poles in the $q_x$ plane and branch points in the $q_z$ plane (the branch in the $q_z$ plane controls the merging of poles in the $q_x$ plane). Another possible branch point in the $q_z$ plane is associated with the square-root in $p$. The fact that a pole in one spectral plane results in a branch point in another spectral plane was recognized in studies of microstrip and other integrated waveguides \cite{IEEEhowto:Chang}-\cite{IEEEhowto:Dennis}. It is also worthwhile to note that the asymptotic methods for branch cut evaluation described in \cite{Felsen} do not work here. To use those formulas the branch cut integral must be dominated by the branch point, that is, by the section of the integral in the vicinity of the branch point. This is not the case for the anisotropic problem, where we have found that sections of the branch cut integral far from the branch point can contribute substantially.

\subsection{p-type branch point in the $q_z-\mathrm{plane}$}
	
For the isotropic case, $ p=\sqrt{q^2-k^2} $ and the p-type branch point occurs at $ q=\pm k $, resulting in the usual hyperbolic branch cuts in the $q$-plane \cite{Ishimaru}. In this case, $ q_x^2+q_z^2=q_p^2 $ is a constant and $ q_z=\sqrt{k^2-q_x^2} $ leads to branch points at $ q_x=\pm k $. However, for the residue, $ q_{p}^2=q^2_{xp}(q_z)+q_z^2 $ is a constant in $ q_z $ and so we never have $q_p = k$ for any $ q_z $, and so there is no p-type BP in the $ q_z-\mathrm{plane} $ for the SPP for the isotropic case. However, for anisotropic media $q_{xp}^{2}\left( q_{z}\right) +q_{z}^{2}$ is
not generally a constant, and so there can be a ``p-type'' BP in the $q_{z}$%
-plane, where $p=\sqrt{q_{xp}^{2}\left( q_{z}\right) +q_{z}^{2}-k^{2}}=0$, although this will not occur at $q_{z}=k$ unless $q_{xp}\left( k\right) =0$. In any event, since this branch cut relates to radiation into space, for the SPP we can ignore this contribution to the SPP field.

Introducing the notation that $ (q^{(n)}_x,q^{(n)}_z) $ represents the pair of spectral values that satisfy the conditions for a branch point/pole pair, (\ref{BP1}) and $\delta \ne 0$, since the residue term already satisfies $D(q_{xp},q_z)=0$, we can find branch points in the $q_z$-plane from $\frac{\partial}{\partial q_{x}}D(q_{xp}(q_z),q_z)= 0 $,

\begin{equation}\label{Eq:25}
\left( \sigma _{xx}+\frac{i k/\eta }{\sqrt{%
		q_{xp}^{2}+q_{z}^{2}-k^{2}}} \left( 1+\frac{1}{4}\eta
^{2}\sigma _{xx}\sigma _{zz}\right) \right)q_{xp}\left( q_{z}\right)\ =0.
\end{equation}
As we will show later, these branch points have a significant role in the analysis of the SPP. Because of their importance, we categorize them into two groups, type-0 and type-1 branch points. 

\subsection{Type-0 branch point in the $q_z$-plane}

First we define type-0 branch points as those values of $ q_z $ for which $ q_{xp}(q_z)=0 $ in (\ref{Eq:25}); i.e., the merging of the forward and backward modes (associated with different signs in the outer square-root in (\ref{Eq:19})) in the $q_x$-plane at a certain value of $q_z$ \cite{1998}, given by 

\begin{equation}\label{Eq:27}
q^{(+\text{0})}=q^{\mathrm{TM}}_z=k \sqrt{1-\left (\frac{2}{\eta \sigma_{zz}} \right )^2}
\end{equation}

\begin{equation}\label{Eq:28}
q^{(-\text{0})} =q^{\mathrm{TE}}_z=k \sqrt{1- \left (\frac{\eta \sigma_{xx}}{2} \right )^2}
\end{equation}

\noindent such that the pair $(q_x,q_z)=(0,q^{\mathrm{TM/TE}})$ form a pole-branch-point pair. For $ \sigma_{xx}=\sigma_{zz} $ these are well-known TM and TE SPP wavenumbers, respectively (graphene is an example of such a 2D isotropic layer which can support these modes \cite{IEEEhowto:Hanson}). Note that for isotropic media, a vertically-polarized current source will produce only TM fields (although a horizontally-polarized source will produce both TE and TM fields even when the sheet is isotropic \cite{Ishimaru}). For an anisotropic sheet the boundary conditions cannot be satisfied assuming only one type of field. 

\subsection{Type-1 branch point in the $q_z$-plane}

Another set of singularities in the  $ q_x$-$q_z$ plane is related to the point in the $q_z$-plane where modes $q_{xp}$ associated with different signs in the inner square-root in (\ref{Eq:19}) merge for $ q_{xp} \neq 0$. These can be obtained by simultaneously solving the equations $ D(q_x,q_z)=0 $ and $ \frac{dD(q_{x},q_z)}{dq_x}=0 $, leading to 

\begin{equation}\label{Eq:29}
q^{(\pm 1)}_x= \sqrt{\frac{-k^2 }{\delta \sigma} \left(\sigma_{xx}+(\sigma_{zz}\mp 2 \sigma_{xx})\frac{(1+\frac{1}{4} \eta^2 \sigma_{xx} \sigma_{zz})^2}{\eta^2 \sigma^2_{xx}}\right)}
\end{equation}

\begin{equation}\label{Eq:30}
q^{(\pm 1)}_z= \sqrt{-(q^{(\pm 1)}_x)^2 +k^2 \left(1-\frac{(1+\frac{1}{4} \eta^2 \sigma_{xx} \sigma_{zz})^2}{\eta^2 \sigma^2_{xx}}\right)}
\end{equation}
where $\delta \sigma=\sigma_{zz}-\sigma_{xx}$, such that $(q_x,q_z)=(q^{(\pm 1)}_x,q^{(\pm 1)}_z)$ form a pole-branch-point pair.

\subsection{Branch cut analysis in the $ q_z$-plane}

Using the SPP field (\ref{Eq:20a}) and performing the outer integration, the Green's function is  

\begin{align}\label{Eq:31}
&g_{\alpha \beta}^{r}= \frac{-i}{2\pi} \int_{-\infty}^{+\infty} w_{\alpha \beta}'(q_{xp},q_z) \frac{e^{-p(y+y')}}{2p} e^{-iq_{xp}(x-x')} \notag \\& ~~~~~~~~ \times e^{-iq_z(z-z')} dq_z. 
\end{align}
Assuming $ (z-z')>0 $, due to the term $ e^{-iq_z(z-z')} $ the contour can be closed in the lower half plane of the $ q_z-\mathrm{plane} $, leading to
\begin{align}\label{Eq:32}
&g_{\alpha \beta}^{r}\approx \frac{-i}{2\pi} \int_{\text{bc}} w_{\alpha \beta}'(q_{xp},q_z) \frac{e^{-p(y+y')}}{2p} e^{-iq_{xp}(x-x')} \notag \\& ~~~~~~~~ \times e^{-iq_z(z-z')} dq_z 
\end{align}
where the branch cut integral is over all branch cuts. Also, from the term $ e^{-iq_{xp}(x-x')} $ it is clear that for $x-x'\gtrless0$ then only when $ \mathrm{Im}(q_{xp})\lessgtr 0 $ do we obtain an SPP that decays away from the source. Therefore, we have in the $ q_z-\mathrm{plane} $ two Riemann sheets (as mentioned previously, neglecting the $p$-type branch point, which would introduce another two sheets; here we simply enforce Re$(p)>0$), the top (proper) sheet where $ \mathrm{Im}(q_{xp})\lessgtr 0 $ and the bottom sheet where $ \mathrm{Im}(q_{xp})\gtrless 0 $, for $x-x'\gtrless0$. Those values of $ q_z $ that lead to $ \mathrm{Im}(q_{xp})=0 $ determine the branch cut trajectory which separates the proper from improper Riemann sheets. 

\begin{figure}[h]
	\begin{center}
		\noindent
		\includegraphics[width=3.5in]{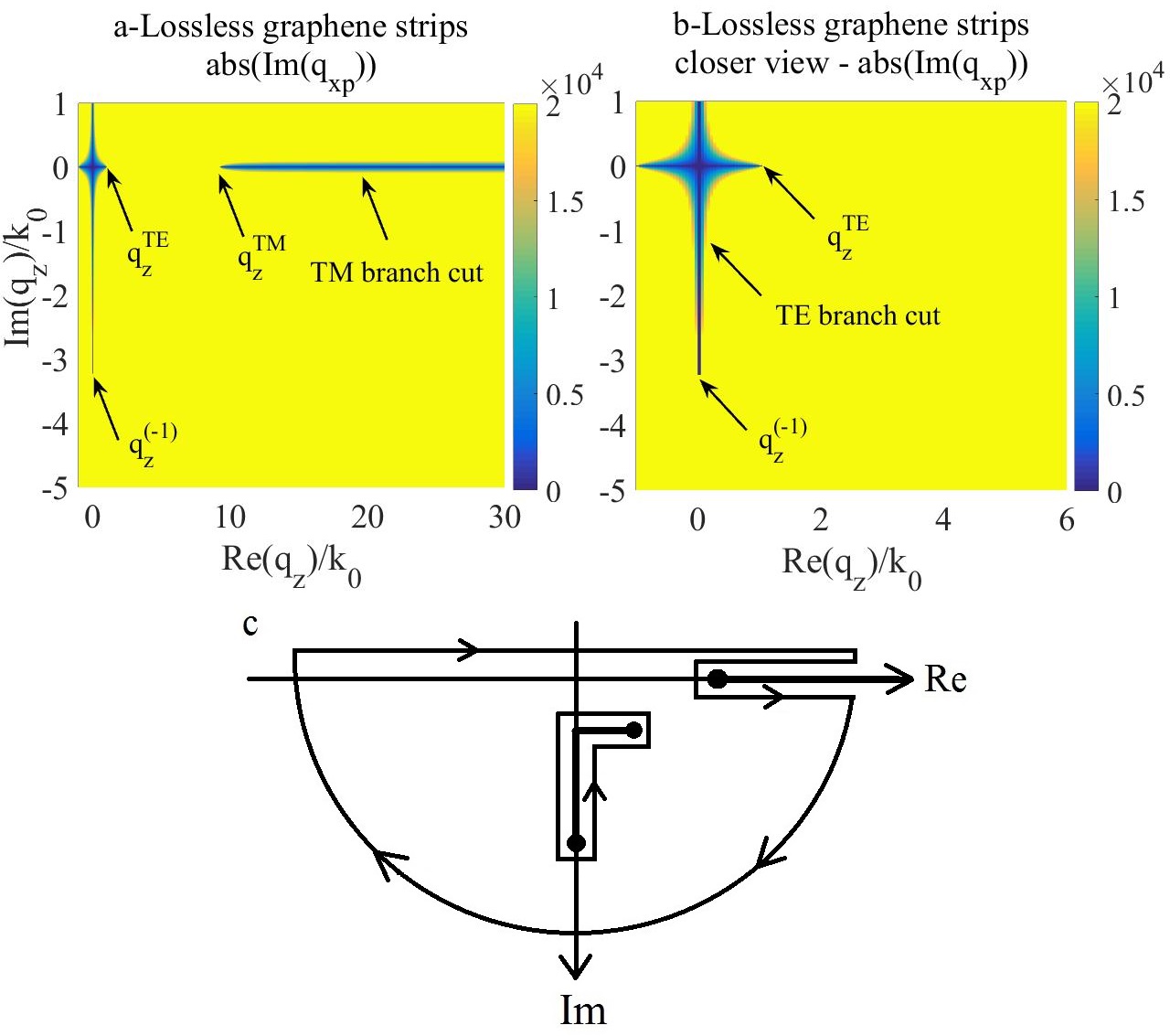}
		\caption{a,b: Branch cut contours  $ \mathrm{Im}(q_{xp})=0 $ determined from a plot of the absolute value of $ \mathrm{Im}(q_{xp}) $ for a lossless model of a graphene strip array at $ 10 $ THz ($ \sigma_{xx}'=\sigma_{zz}'=0 $, $ \sigma_{xx}^{\prime \prime}= 0.57i $ mS, $ \sigma_{zz}^{\prime \prime}= - 0.57i$ mS). The branch point locations are $q^{\mathrm{TE}}_z/k=1.005$, $q^{\mathrm{TM}}_z/k=9.3$, $q^{(-1)}_z/k=-3.22i$. c. Integration contour in the $ q_z-\mathrm{plane} $ showing branch points (dots) and branch cuts (thick lines). }\label{Fig3}
	\end{center}
\end{figure}

Typically, branch cut trajectories to separate certain Riemann sheets can be analytically determined from the functional dependence of the multi-valued function that defines the branch point. However, for anisotropic surfaces the form of $ q_{xp} $ is too complicated to determine a simple equation for the branch cut for $ \mathrm{Im}(q_{xp})= 0 $. As an example, Fig. \ref{Fig3}-a shows the branch cuts for $ \mathrm{Im}(q_{xp})=0 $ obtained by plotting Im($q_{xp}$) for an array of graphene strips (see Appendix A) in the hypothetical lossless case (i.e., ignoring the real parts of the conductivities) at $ 10 $ THz. Fig. \ref{Fig3}-b shows a close-up near the Im axis, and Fig. \ref{Fig3}-c shows the properly-cut $ q_z-\mathrm{plane} $ for the lossless case. It can be seen that for the considered frequency the TM branch point leads to a branch cut starting at $q^{\mathrm{TM}}_z$ and going horizontally to infinity, and the TE branch point $q^{\mathrm{TE}}_z$ and the branch point $q^{(-1)}_z$ are connected by a branch cut. The branch point $q^{(+1)}_z$ is on the improper Riemann sheet (not shown). 

Insight into the correct branch cut can be obtained from a large $ q_z $ approximation. From (\ref{Eq:21}), for a lossy 2D surface $ \sigma_{xx}=\sigma_{xx}'+i \sigma_{xx}^{\prime \prime} $ and $ \sigma_{zz}=\sigma_{zz}'+i \sigma_{zz}^{\prime \prime} $ then the branch cut trajectory is along $ q_z $ values such that 

\begin{equation}\label{Eq:33}
\mathrm{Im}(iq_z \sqrt{\sigma_{zz}'\sigma_{xx}'+i\sigma_{zz}^{\prime \prime}\sigma_{xx}'-i\sigma_{xx}^{\prime \prime}\sigma_{zz}'+\sigma_{xx}^{\prime \prime}\sigma_{zz}^{\prime \prime}}) = 0.
\end{equation}

\noindent For a lossless surface, $ \sigma_{xx}'=\sigma_{zz}'=0 $, leading to

\begin{equation}\label{Eq:34}
\mathrm{Im}(iq_z \sqrt{\sigma_{xx}^{\prime \prime}\sigma_{zz}^{\prime \prime}}) = 0, 
\end{equation}
such that if $\sigma_{xx}^{\prime \prime}\sigma_{zz}^{\prime \prime}>0$ the BC is along $ \mathrm{Im}(q_z) $, and if $\sigma_{xx}^{\prime \prime}\sigma_{zz}^{\prime \prime}<0$ the BC is along $ \mathrm{Re}(q_z) $, in agreement with the numerically-determined contours.

The branch cut integrals can be viewed as a continuous superposition of modes. The BP  $ q_z^{\mathrm{TM}} $ is associated with the pair $(q_x,q_z)=(0, q_z^{\mathrm{TM}} ) =(0,9.3)k$ for the numerical example considered), and along the branch cut, as Re($q_z$) increases, Re($q_{x}$)=Re($q_{xp}$) also increases from zero, and the resulting continuum summation of pair values synthesis the beam. Similar comments apply to the branch cut between $ q_z^{\mathrm{TE}} $ and $ q_z^{-1}$ (between $q_z=1.005k$ and $-3.22ik$ in the numerical example considered).

\begin{figure}[h]
	\begin{center}
		\noindent
		\includegraphics[width=3.5in]{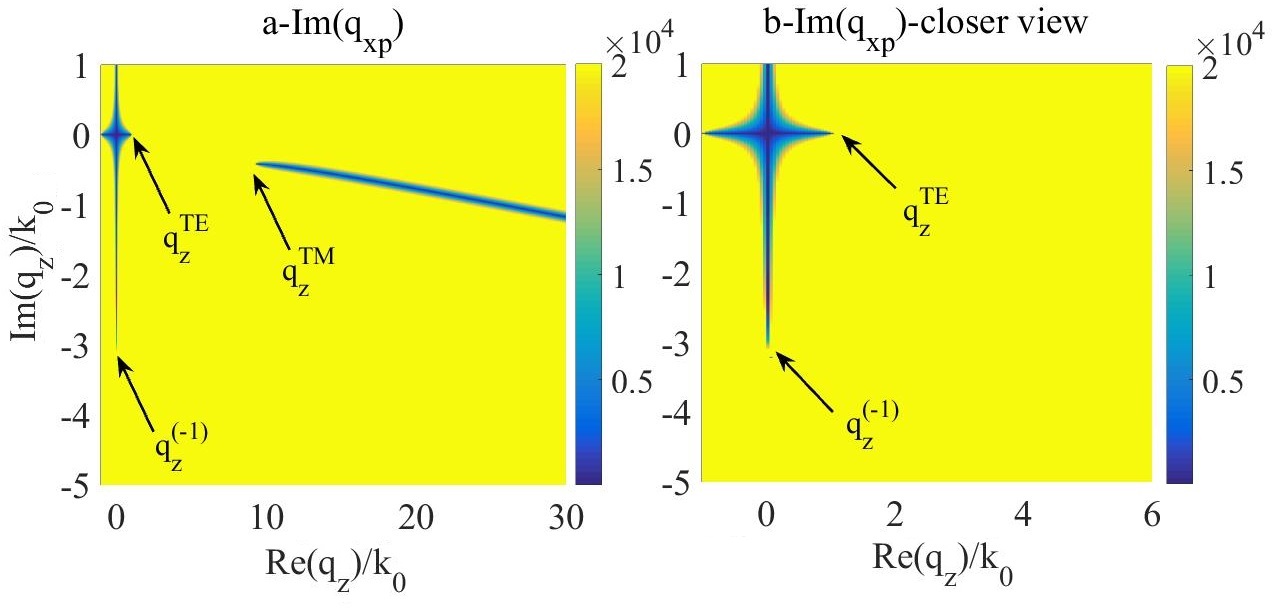}
		\caption{Branch cut contours $ \mathrm{Im}(q_{xp})=0 $ determined from a plot of the absolute value of $ \mathrm{Im}(q_{xp}) $ for a lossy model of a graphene strip array at $ 10 $ THz with $ \sigma_{xx}=0.02 + 0.57i $ mS and $ \sigma_{zz} = 0.02 - 0.57i $ mS.}\label{Fig4}
	\end{center}
\end{figure}

The lossy case is shown in Fig. \ref{Fig4}; the branch cut trajectory deflects a bit from the lossless case, but for low-loss surface the lossless BC contour is sufficient. 

\begin{figure}[h]
	\begin{center}
		\noindent
		\includegraphics[width=2.5in]{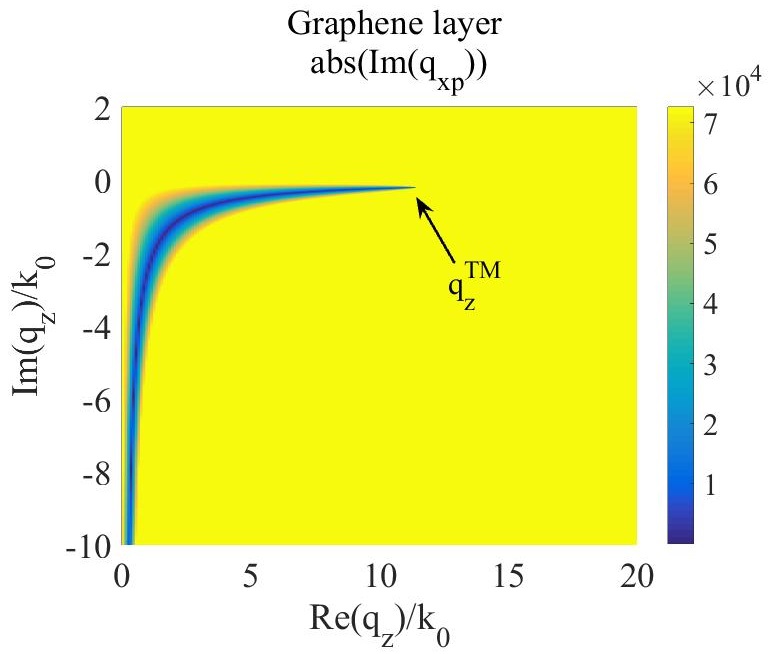}
		\caption{Branch cut contour $ \mathrm{Im}(q_{xp})=0 $ determined from a plot of the absolute value of $ \mathrm{Im}(q_{xp}) $ for graphene with $ \mu_c=0.5$ eV at $ T=0$ K and $ f=20 $ THz.}\label{Fig5}
	\end{center}
\end{figure}

As a common special case, for an inductive isotropic surface such as graphene in the far-infrared, 

\begin{align}
\sigma_{xx} =\sigma_{zz} = & \frac{-ie^2k_BT}{\pi\hbar^2(\omega-i2\Gamma)} \times \notag \\
&\left(\frac{\mu_c}{k_BT}+2\ln\left(1+e^{-\frac{\mu_c}{k_BT}}\right)\right). \label{Eq:35}
\end{align}
Here we consider graphene at $ T=300 $ K, $ \mu_c=0.5$ eV and $ f=20 $ THz. In this case the TE related branch point is at $ q_z^{\mathrm{TE}}=k(1.0039 + 0.0001i) $, and so is not implicated in the lower-half-plane closure, consistent with the surface being inductive (no TE mode is supported). Since only TM branch points occur, only a TM mode exists, and the TM related BP occurs at $ q_z^{\mathrm{TM}}/k=(11.3706 - 0.2088i) $. The two other type-1 branch points move to infinity as the surface becomes isotropic, and therefore the branch cut extends down the entire imaginary axis (therefore for both the isotropic and anisotropic cases there is a branch cut between $ q_z^{\mathrm{TM}} $ and $ q_z^{-1} $). Fig. \ref{Fig5} shows a surface plot of $ \mathrm{Im}(q_{xp}) $ in the $ q_z-\mathrm{plane} $.

For isotropic and inductive graphene only a TM mode can propagate, and so the contribution is from the TM-related branch point and associated cut, as expected. For the graphene strip array anisotropic case, the hybrid nature of the modes supported by such a surface involve both TE and TM-related branch points, and, in contrast to the isotropic case, three branch points contribute to the field. 

\subsection{Conductivity and its effect on branch points and SPP confinement}

Analytically it can be shown that both type-1 branch points $ q_z^{(\pm 1)} $ can be connected to a TE or TM branch point, depending on the conductivity value. Two cases are of particular interest,  small conductivity values, $ (\mathrm{Im}(\sigma_{xx/zz})\eta)^2 \ll 1$, and large conductivity values,  $ ( \mathrm{Im}(\sigma_{xx/zz})\eta)^2 \gg 1$. For small conductivity values, from (\ref{Eq:27})-(\ref{Eq:28}) we have 

\begin{equation}\label{Eq:36}
q^{TM}_z=k \sqrt{1-\left (\frac{2}{\eta \sigma_{zz}} \right )^2}~ \longrightarrow ~ (\eta \sigma_{zz})^2=\frac{4}{1-(\frac{q_z^{\mathrm{TM}}}{k})^2}
\end{equation}

\begin{equation}\label{Eq:37}
q^{TE}_z=k \sqrt{1- \left (\frac{\eta \sigma_{xx}}{2} \right )^2} ~ \longrightarrow ~ \frac{1}{(\eta \sigma_{xx})^2}=\frac{1}{4} \frac{1}{1-(\frac{q_z^{\mathrm{TE}}}{k})^2}.
\end{equation} 
Making these replacements in (\ref{Eq:29})-(\ref{Eq:30}) and using the fact that for small conductivity like in our previous numeric example ($\sigma_{xx}= 0.02 + 0.57i$ mS and $\sigma_{zz} = 0.02 - 0.57i$ mS) we have $ (\mathrm{Im}(\sigma_{xx/zz})\eta )^2 \ll 1 $, then $ |q_z^{\mathrm{TM}}|\gg k $ and $ |q_z^{\mathrm{TE}}| \approx k$, and so $ |q_z^{\mathrm{TE}}|^2 \ll |q_z^{\mathrm{TM}}|^2  $, such that  

\begin{equation}\label{Eq:38}
q_z^{(\pm 1)}= \frac{k}{2} \sqrt{\frac{1}{1-(\frac{q_z^{\mathrm{TE}}}{k})^2} \frac{\sigma_{xx}\mp 2\sigma_{xx}}{\sigma_{zz}-\sigma_{xx}}}.
\end{equation}
Therefore, for small values of $\sigma_{xx}$ and $\sigma_{zz}$, the type-1 branch points are governed by (and associated with) the TE branch point $ q_z^{\mathrm{TE}} $. 

For larger values of $\sigma_{xx}$ and $\sigma_{zz}$ the situation is different. In this case, for $ (\mathrm{Im}(\sigma_{xx/zz})\eta )^2 \gg 1$ we have $|q_z^{\mathrm{TM}}|^2 \ll |q_z^{\mathrm{TE}}|^2 $ and it can be shown that an approximate expression for the type-1 branch point is (\ref{Eq:38}) with $q_z^{\mathrm{TM}}$ replacing $q_z^{\mathrm{TE}}$; the type-1 branch points are associated with the TM related branch point. As the conductivity changes from a small to a large value, $ q_z^{TE} $ and $ q_z^{TM} $ move toward each other and then cross, and eventually interchange roles. Setting (\ref{Eq:27}) and (\ref{Eq:28}) equal to each other, it can be shown that these type-0 branch points meet at a frequency such that $ \sigma_{xx} \sigma_{zz}=4/\eta^2 $.

\begin{figure}[h]
	\begin{center}
		\noindent
		\includegraphics[width=3.5in]{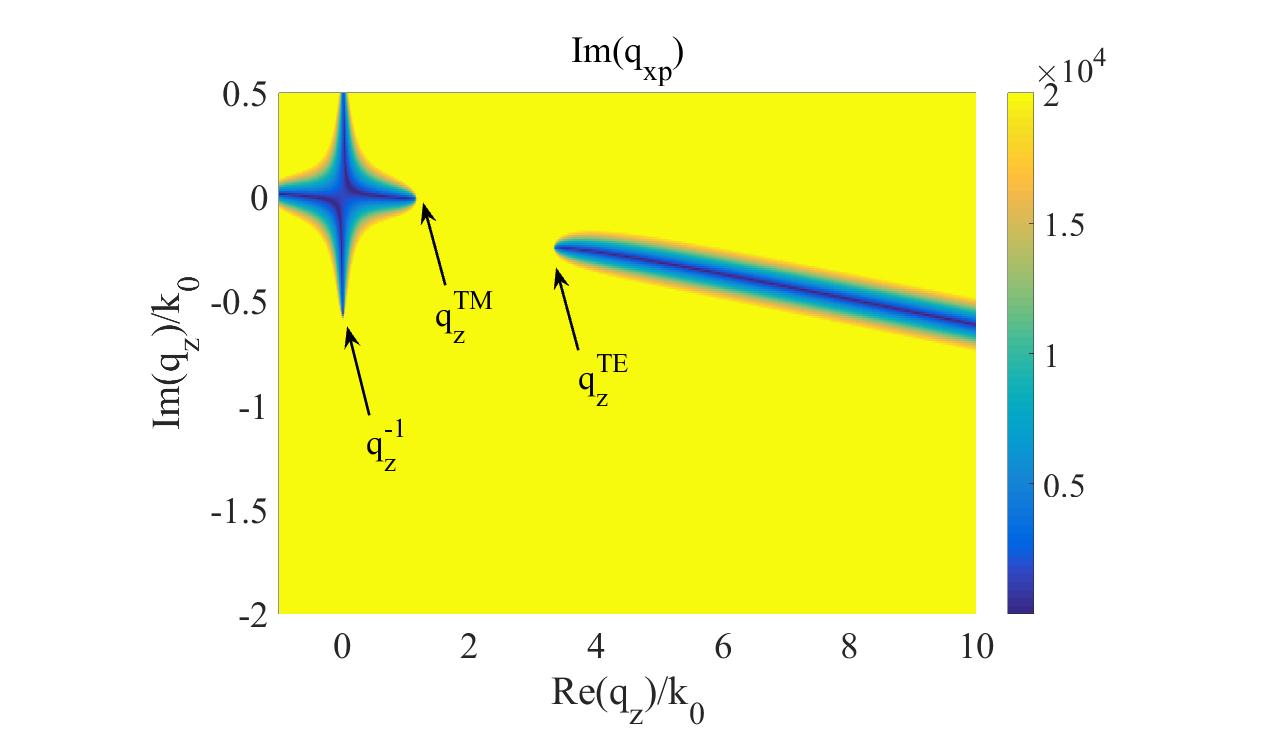}
		\caption{Branch cut contours $ \mathrm{Im}(q_{xp})=0 $ determined from a plot of the absolute value of $ \mathrm{Im}(q_{xp}) $ for a lossy model of multi-layer graphene strip at $ 10 $ THz, $ \sigma_{xx} = 1.3 +16.9i  $ mS and $ \sigma_{zz} = 0.4 - 9.2i  $ mS.}\label{Fig7}
	\end{center}
\end{figure}

As an example of a large conductivity situation, conductivity tensor components $ \sigma_{xx} = 1.3 +16.9i  $ mS and $ \sigma_{zz} = 0.4 - 9.2i  $ mS are attainable using multi-layer graphene to form the strip array. For this set of conductivities the branch points and branch cuts are shown in Fig. \ref{Fig7}. As can be seen, $ q_z^{TE} $ exceeds $ q_z^{TM} $, there is a branch cut from $ q_z^{TE} $ to infinity, a branch cut between $ q_z^{TM} $ and $ q_z^{-1} $, and $ q_z^{-1} $ is connected to $ q_z^{TM} $.

\section{Directive SPPs on Hyperbolic and Anisotropic Surfaces} \label{MR}

\subsection{Anisotropic hyperbolic  layer (graphene strip array)}
\begin{figure}[h]
	\begin{center}
		\noindent
		\includegraphics[width=3.5in]{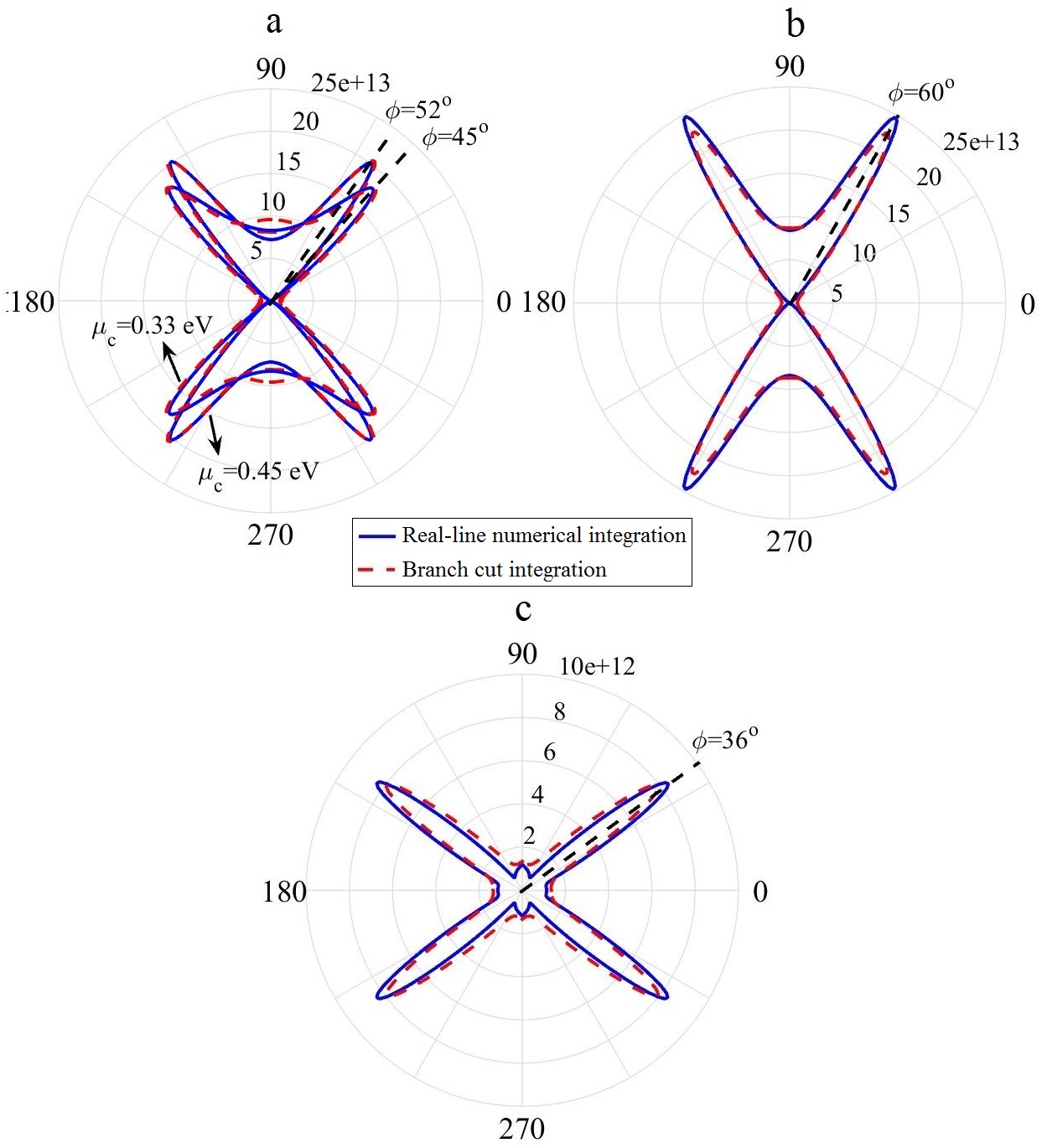}
		\caption{Electric field $ E_{y} $ excited by a $y$-directed dipole current above a graphene strip array. a: graphene with $ \mu_c=0.45 $ eV and $ \mu_c= 0.33 $ eV, $ W=59$ nm, $ L=64$ nm, $ \sigma_{xx} = 0.02 + 0.57i $ mS and $ \sigma_{zz} = 0.02 - 0.57i $ mS. b: $ \mu_c= 0.45 $ eV, $ W=56.1$ nm, $ L=62.4$ nm, $ \sigma_{xx} =0.003 + 0.25i$ mS and $ \sigma_{zz} =0.03 - 0.76i$ mS. c: Strip array with a 5-layer graphene, $ \mu_c= 1 $ eV, $ W=196$ nm and $ L=200$ nm, $ \sigma_{xx} = 1.3 +16.9i  $ mS and $ \sigma_{zz} = 0.4 - 9.2i  $ mS. Blue line is for the integration along the real axis (\ref{Eq:31}) and dashed red line is for integration along the branch cuts (\ref{Eq:32}). $f=10$ THz, $\rho=0.2\lambda$, and $y=0.005\lambda$.}\label{Fig8}
	\end{center}
\end{figure}

As shown in Appendix A, conductivity components $ \sigma_{xx} = 0.02 + 0.57i $ mS and $ \sigma_{zz} = 0.02 - 0.57i $ mS can be realized using an array of graphene strips with $ \mu_c= 0.33$ eV, strip width $ W=59$ nm, and period $ L=64$ nm. For this anisotropic hyperbolic surface, Fig. \ref{Fig8}a shows the electric field  $ E_{y} $, the dominant field component, computed as a real-line integral (\ref{Eq:31}), and as a sum of branch cut integrals (\ref{Eq:32});  excellent agreement is found between the two methods (the branch cut integrals are faster to compute than the brute-force numerical integrals, but no attempt was made to optimize either integration). The branch cuts for this case are shown in Fig. \ref{Fig4}. Figs. \ref{Fig8}-b,c show similar agreement for different strip configurations as discussed below.

Although the direction of the beam is electronically controllable via the chemical potential, different combinations of physical parameters of the graphene strip array (width $W$ and periodicity $L$) can also be used to produce a desired beam. An optimum geometry to produce a beam in a certain direction can be found by tuning all of these parameters simultaneously. 

From (\ref{Eq:22}), in the hyperbolic regime propagation along a desired direction can be obtained if the tensor conductivity components have the proper ratio. Designing a hyperbolic metasurface to produce a beam in a desired direction (e.g., choosing the strip width and period) can be done by trial-and-error tuning of all geometrical and electrical parameters of the system, but a multi-variable optimization, such as a genetic algorithm (GA) is a good choice for this task \cite{Randy} \cite{Randy-book}. Ideally, the physical layout of the metasurface (graphene strips in the case) should be designed so that the effective (homogenized) conductivity tensor elements are hyperbolic, and have large imaginary part and small real part, since such a surface can support a well-confined, long-range SPP. here we used the cost function to be minimized as

\begin{align}\label{Eq:41}
	\Psi(L,W,&\mu_c,\phi) = \alpha (\mathrm{Re}(\sigma_{xx})+\mathrm{Re}(\sigma_{zz})) \notag \\& 
	+ \frac{\beta}{| \mathrm{Im}(\sigma_{xx}) |+| \mathrm{Im}(\sigma_{zz})|}  +\gamma\left(\tan^2(\phi)+ \frac{\sigma_{zz}}{\sigma_{xx}}\right)
\end{align}
where $\sigma_{xx}$ and $\sigma_{zz}$ are defined in (\ref{Eq:40}) in Appendix A. The cost function in (\ref{Eq:41}) is a multi-objective cost function and the coefficients $\alpha$, $\beta$ and $\gamma$ assign a weight (0 to 1) to each objective regarding to its importance. The first term in (\ref{Eq:41}) assures a small real part of conductivity, the second term assures a large imaginary part, and the last term assures the correct ratio for $ \sigma_{zz} $ and $ \sigma_{xx} $ to obtain the SPP beam in desired direction specified by $ \phi $. It was found that $ \alpha=0.2 $ and $\beta=\gamma = 0.4$ leads to good results.

The physical strip geometry leading to the beam in Fig. \ref{Fig8}-a was found in this manner, for a specified beam angle of 45 deg. Note the excellent agreement between desired and obtained beam angle. The chemical potential was then changed to produce the beam at 52 deg., for a fixed geometry. Thus, a significant aspect of using a graphene strip array is its electronic tunability by, e.g., varying the bias to control the chemical potential.

In Fig. \ref{Fig8}-b a desired beam angle of 60 deg. was sought, and the GA was used to determine the optimized parameters; $ \mu_c= 0.45 $ eV, $ W=56.1$ nm and $ L=62.4$ nm, such that $ \sigma_{xx} =0.003 + 0.25i$ mS  and $ \sigma_{zz} =0.03 - 0.76i$ mS, leading to the desired beam. Again, excellent agreement is found between the desired and final beam angles.  

As a final example for the graphene strip array Fig. \ref{Fig8}-c shows $ E_y $ for the case of multi-layer graphene strips (to increase the conductivity) as discussed in the previous section. By using five layers of graphene with $ \mu_c= 1 $ eV, $ W=196$ nm and $ L=200$ nm, the conductivities are $ \sigma_{xx} = 1.3 +16.9i  $ mS and $ \sigma_{zz} = 0.4 - 9.2i  $ mS. The branch cuts are shown in Fig. \ref{Fig7}. For this case, (\ref{Eq:22}) indicates that the beam should be directed along $ \phi=36 $ deg. Again, excellent agreement is found between the two methods and the position of the beam is along the desired angle.

\subsection{Anisotropic non-hyperbolic layer (black phosphorus)}

As discussed in Appendix C, black phosphorus is a natural material that can be used as a platform to realize an anisotropic surface. Although black phosphorus exhibits a hyperbolic regime, the resulting values of conductivity are rather small (to produce a hyperbolic response the interband conductivity must dominate one of the conductivity values ($\sigma_{xx}$ or $\sigma_{zz}$), and the intraband conductivity must dominate the other component, resulting in the required sign difference). Although a hyperbolic SPP can be excited, the residue is not generally the dominant response. Therefore, in order to consider larger values of black phosphorus conductivity, we consider the non-hyperbolic (Drude) regime. A $ 10 $ nm thick black phosphorus film with doping level $ 10\times10^{13} $/cm$^2$ has conductivity tensor components $ \sigma_{xx} = 0.0008 - 0.2923i  $ mS and $ \sigma_{zz} =  0.0002 - 0.0658i $ mS at $ f=92.6 $ THz. Using (\ref{Eq:27}), (\ref{Eq:28}) and (\ref{Eq:30}), a surface with these conductivity components has $q_z^{\mathrm{TM}} = k(80.6804 - 0.2114i) $, $ q_z^{\mathrm{TE}} \approx k $, and $ q_z^{(-1)} = k(-0.0300 -10.3165i) $.

\begin{figure}[h]
	\begin{center}
		\noindent
		\includegraphics[width=3.5in]{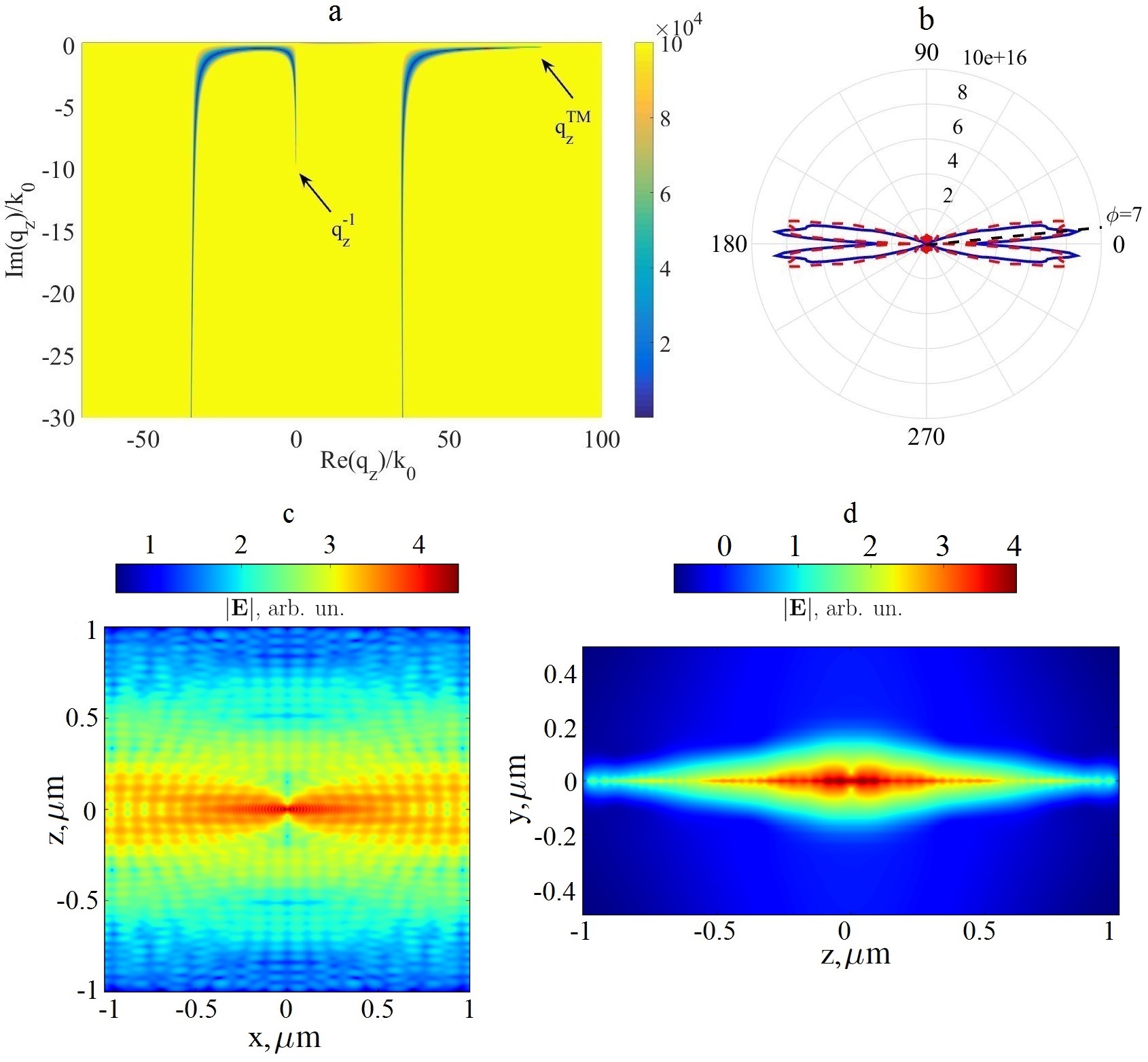}
		\caption {a: Branch cut contours for $\mathrm{Im}(q_{xp})=0$  determined from the absolute value of $\mathrm{Im}(q_{xp}) $ in the $ q_z-\mathrm{plane} $. b. Absolute value of $ E_{y} $ excited by a $y$-directed dipole current source above black phosphorus with doping level $ 10\times10^{13} $/cm$^2$ at $ f=92.6 $ THz. The blue line is for the integration along the real axis (\ref{Eq:31}) and the dashed red line is for the  integration along the branch cuts (\ref{Eq:32}). $\rho=0.2\lambda$ and $y=0.005\lambda$. c: SPP field in-plane distribution in logarithmic scale calculated by FDTD. d: SPP field vertical variation in logarithmic scale calculated by FDTD.}\label{Fig9}
	\end{center}
\end{figure}

The imaginary components of the conductivities are negative, so that the surface is not able to support TE modes (the TE branch point is located at the upper half of the $ q_z-\mathrm{plane} $, and so not captured for $z-z'>0$). The only active branch points are the TM related branch point and $ q_z^{(-1)} $. Fig. \ref{Fig9}-a shows the branch points and associated branch cuts in the $ q_z-\mathrm{plane} $. One important difference between branch cuts in this case and in the previous hyperbolic cases is the branch cut trajectory. From (\ref{Eq:34}) for the hyperbolic case, because of the condition $ \mathrm{Im}(\sigma_{xx}) \mathrm{Im} (\sigma_{zz}) <0 $ the branch cut trajectory was along the real axis, but for the anisotropic non-hyperbolic case we have $ \mathrm{Im}(\sigma_{xx}) \mathrm{Im} (\sigma_{zz}) >0 $ and so the trajectory for large $ q_z $ is parallel to the imaginary axis.

As shown in Fig. \ref{Fig9}-b, this anisotropic non-hyperbolic surface can support a directed SPP, although the beam is directed primarily along one of the coordinate axes. The electric field computed as a real-line integral (\ref{Eq:31}) is in good agreement with the electric field obtained as a sum of branch cut integrals (\ref{Eq:32}). Fig. \ref{Fig9}-c shows the SPP field in logarithmic scale calculated by numerically solving Maxwell's equations using a commercial finite-difference time-domain method (FDTD) from Lumerical solutions \cite{IEEEhowto:Lumerical}. Good agreement with the results obtained by complex plane analysis is observed. Fig. \ref{Fig9}-d shows the vertical variation of the beam in logarithmic scale calculated by Lumerical, showing strong SPP confinement to the surface. Using the Green's function the attenuation length was found to be $p=\lambda / 12\pi$.

\section{Conclusion}

We have studied the electromagnetic response of two-dimensional anisotropic and hyperbolic surfaces and developed a method (based on complex plane analysis) for the efficient computation of electric field excited on such surfaces. A solution in term of electric field Sommerfeld integrals has been obtained for the electromagnetic field due to a vertical dipole current source located in close proximity to the surface. Poles, branch points, and related branch cuts and their relative importance and physical meaning for surface wave propagation has been emphasized. A first-order approximation has also been obtained using the stationary phase method. Examples have been shown for a graphene strip array and black phosphorus.

\appendices
\section{Graphene strip hyperbolic metasurface}

\begin{figure}[h]
	\begin{center}
		\noindent
		\includegraphics[width=3.5in]{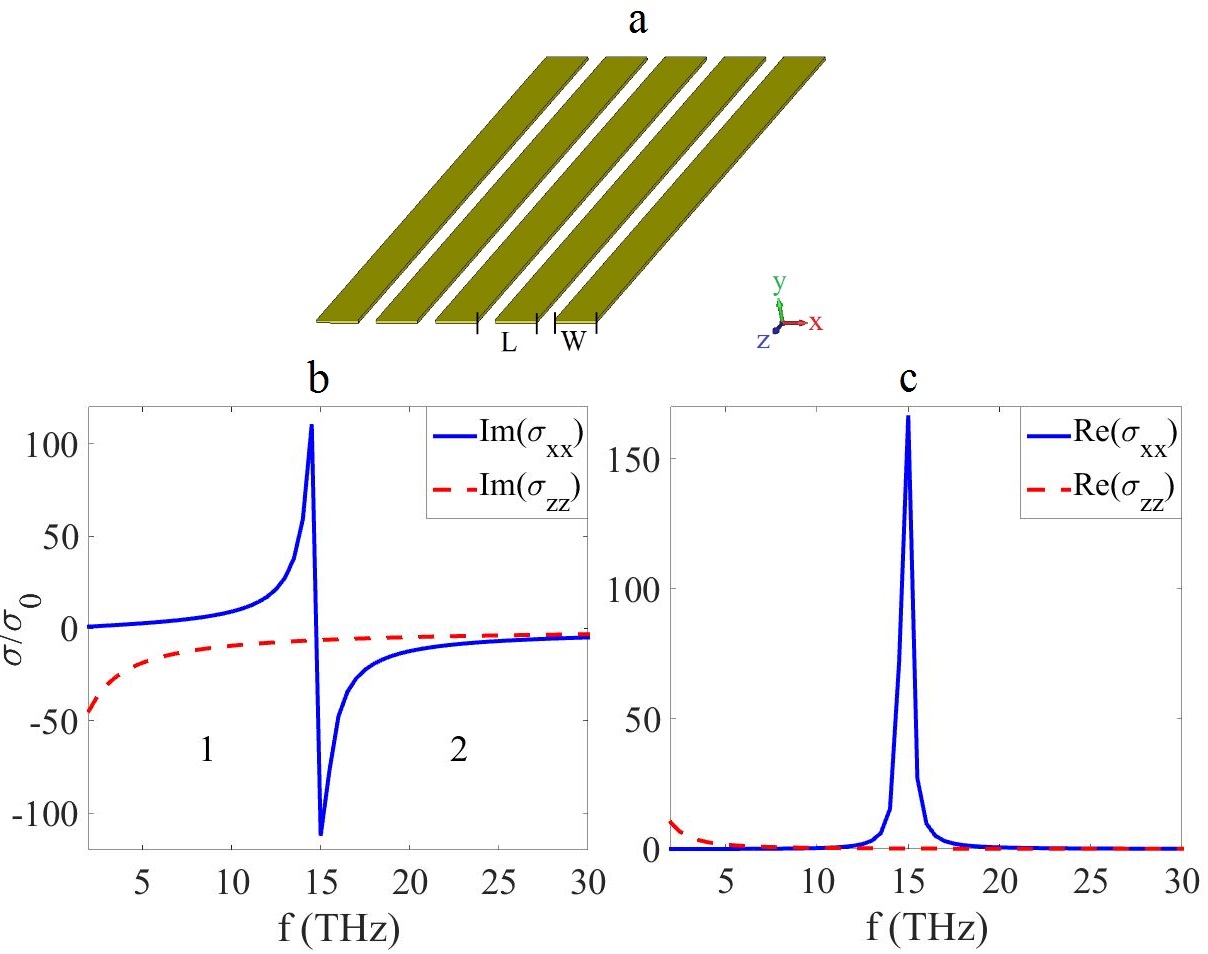}
		\caption{a: Array of graphene strips. b: Imaginary parts of $ \sigma_{xx} $ and $ \sigma_{zz} $ and c: real parts of $ \sigma_{xx} $ and $ \sigma_{zz} $ normalized to $ \sigma_0 = e^2/4 \hbar $ for a graphene strip array with $ \tau = 0.35 $ ps, $ \mu_c=0.33 $ eV, $ W=59 $ nm and $ L = 64 $ nm. Region 1 is hyperbolic and region 2 is simply anisotropic}\label{Fig10}
	\end{center}
\end{figure}

A schematic of an array of graphene strips is shown in Fig. \ref{Fig10}-a. This densely packed strip surface can act as a physical implementation of a metasurface at terahertz and near infrared frequencies \cite{IEEEhowto:Alu,Shapoval}. The dispersion topology of the proposed structure may range from elliptical to hyperbolic as a function of its geometrical and electrical parameters. The in-plane effective
conductivity tensor of the proposed structure can be analytically obtained using an effective medium theory as \cite{IEEEhowto:Alu}

\begin{equation}\label{Eq:40}
\sigma_{zz}^{\mathrm{eff}}=\sigma \frac{W}{L} ~~ \mathrm{and} ~~ \sigma_{xx}^{\mathrm{eff}}=\frac{L \sigma \sigma_c}{W \sigma_c + G \sigma}
\end{equation}
where $L$ and $W$ are the periodicity and width of the strips, respectively, $ G=L-W $ is the separation distance between two consecutive strips, $\sigma $ is graphene conductivity (\ref{Eq:35}) and $ \sigma_c=j\frac{\omega \epsilon_0 L}{\pi} \text{ln} (\text{csc} \frac{\pi G}{2L}) $ is an equivalent conductivity associated with the near-field coupling between adjacent strips obtained using an electrostatic approach \cite{Kaipa}. These effective parameters are valid only when the homogeneity
condition $ L\ll \lambda_{\mathrm{SPP}} $ is satisfied, where $ \lambda_{\mathrm{SPP}} $ is the plasmon wavelength in the in-plane direction perpendicular to the strips ($ x $ in this case), thus leading to a homogeneous 2D metasurface. Fig. \ref{Fig10}-b and c shows $ \sigma_{xx} $ and $ \sigma_{zz} $ in a wide range of frequency for a graphene strip array with graphene parameters $ \tau = 0.35 $ ps, $ \mu_c=0.33 $ eV, and geometrical parameters $ W=59 $ nm and $ L = 64 $ nm. As can be seen from in Fig. \ref{Fig10}-b, this structure can exhibit a hyperbolic response, as well as implement a non-hyperbolic although anisotropic surface. 

\section{Black phosphorus}

\begin{figure}[h]
	\begin{center}
		\noindent
		\includegraphics[width=3.6in]{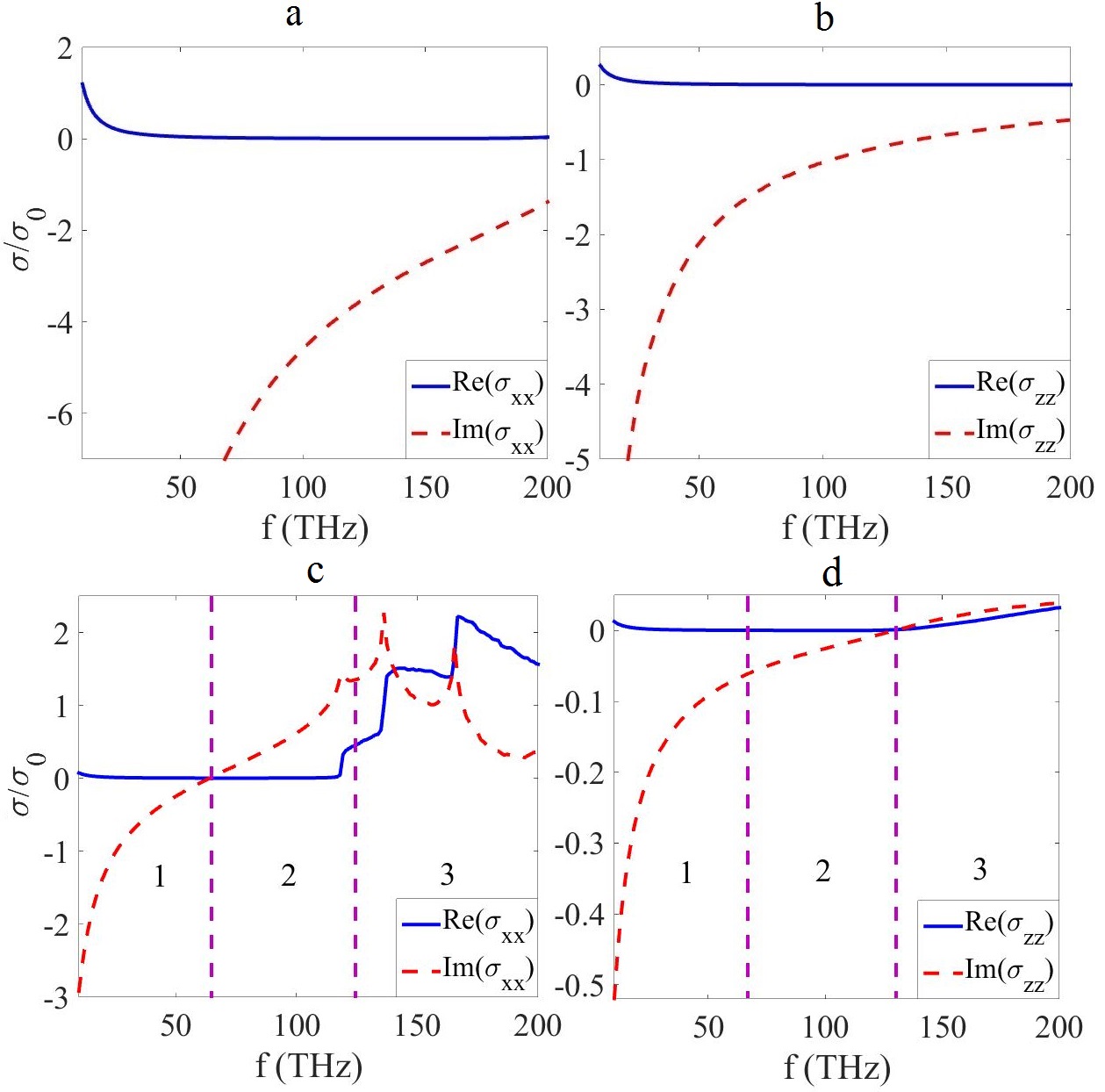}
		\caption{a, b. Real and imaginary parts of $ \sigma_{xx} $ and $ \sigma_{zz} $ ($x$ and $z$ are in-plane crystal axes of BP, with $x$ along the small effective mass direction, or commonly called the armchair direction) obtained at doping level $ 10\times10^{13} $/cm$^2$ and c,d. $ 5\times10^{12} $/cm$^2$ normalized to $ \sigma_0 =  e^2/ 4 \hbar $ with a 10 nm thickness. Regions 1 and 3 show anisotropic inductive and capacitive responses, respectively, and region 2 shows the hyperbolic regime. T=300 K and damping is 2 meV.}\label{Fig11}
	\end{center}
\end{figure}

Black phosphorous is an anisotropic monolayer or thin-film material that can support surface plasmons \cite{IEEEhowto:Low}. Fig. \ref{Fig11} shows the in-plane conductivity tensor components at two doping levels, $ 10\times10^{13} $/cm$^2$ in Fig. \ref{Fig11}-a and b and $ 5\times10^{12} $/cm$^2$ in Fig. \ref{Fig11}-c and d, obtained from a Kubo formula as described in \cite{IEEEhowto:Tony3}. For a 10 nm BP film, the electronic band gap is approximately 0.5 eV. This accounts for the observed interband absorption along the x polarization, and also characterized by weak interband absorption along z.

It can be seen that by increasing the doping level, larger conductivity components are attainable but the hyperbolic region is also pushed toward higher frequencies. In Fig. \ref{Fig11}-a and b black phosphorus is an inductive anisotropic (non-hyperbolic) surface while in Fig. \ref{Fig11}-c and d regions 1 and 3 show anisotropic inductive and capacitive responses, respectively, and region 2 shows the anisotropic hyperbolic region. 

\section*{Acknowledgment}

The authors would like to thank the reviewers for their careful consideration and suggestions. TL was supported partially by the MRSEC Program of the National Science Foundation under Award Number DMR-1420013

\end{document}